\documentclass[journal=jacsat,manuscript=article]{achemso}

\usepackage[utf8]{inputenc}
\usepackage{textcomp}
\DeclareUnicodeCharacter{2212}{\textminus}
\usepackage[version=3]{mhchem} % Formula subscripts using \ce{}
\usepackage{amsmath}
\usepackage{soul}
\author{Celestine Lalengmawia}
\affiliation[MZU]
{Department of Physics, Mizoram University, Aizawl-796004, India}
\alsoaffiliation[PUC]
{Physical Sciences Research Center (PSRC), Department of Physics, Pachhunga University College, Mizoram University, Aizawl-796001, India}

\author{Michael T. Nunsanga}
\affiliation[NITM]
{Department of Mechanical Engineering, National Institute of Technology (NIT) Mizoram, Chaltlang, Aizawl-796012, India.}

\author{Saurav Suman}
\affiliation[NITM]
{Department of Mechanical Engineering, National Institute of Technology (NIT) Mizoram, Chaltlang, Aizawl-796012, India.}

\author{Zosiamliana Renthlei}
\affiliation[MZU]
{Department of Physics, Mizoram University, Aizawl-796004, India}
\alsoaffiliation[PUC]
{Physical Sciences Research Center (PSRC), Department of Physics, Pachhunga University College, Mizoram University, Aizawl-796001, India}

\author{Lalruat Sanga}
\affiliation[MZU]
{Department of Physics, Mizoram University, Aizawl-796004, India}
\alsoaffiliation[PUC]
{Physical Sciences Research Center (PSRC), Department of Physics, Pachhunga University College, Mizoram University, Aizawl-796001, India}
\author{Hani Laltlanmawii}
\affiliation[MZU]
{Department of Physics, Mizoram University, Aizawl-796004, India}
\alsoaffiliation[PUC]
{Physical Sciences Research Center (PSRC), Department of Physics, Pachhunga University College, Mizoram University, Aizawl-796001, India}

\author{Lalhriat Zuala}
\affiliation[PUC]
{Physical Sciences Research Center (PSRC), Department of Physics, Pachhunga University College, Mizoram University, Aizawl-796001, India}

\author{Shivraj Gurung}
\affiliation[PUC]
{Physical Sciences Research Center (PSRC), Department of Physics, Pachhunga University College, Mizoram University, Aizawl-796001, India}

\author{Amel Laref}
\affiliation{Department of Physics and Astronomy, College of Science, King Saud University, Riyadh, 11451, Saudi Arabia}

\author{Dibya Prakash Rai}
\email{dibyaprakashrai@gmail.com}
\affiliation[MZU]
{Department of Physics, Mizoram University, Aizawl-796004, India}

\title[An \textsf{achemso} demo]
  {Exploring the functional properties of diamond-like quaternary compound- Li$_2$ZnGeS$_4$ for potential energy applications: A theoretical approach}

\abbreviations{IR,NMR,UV}

\begin{document}

\begin{tocentry}
\includegraphics[width=8.5cm, height=3.5cm]{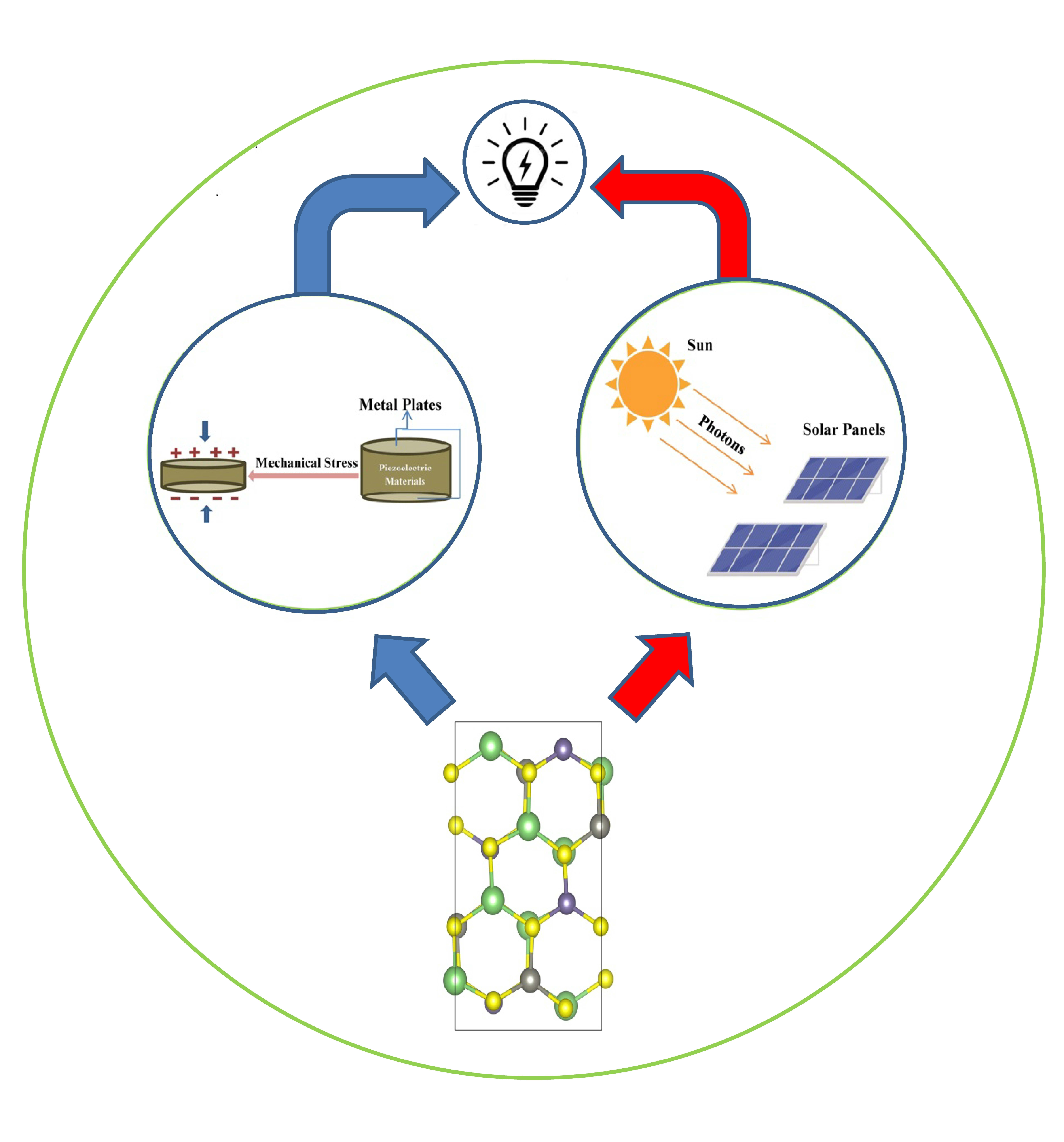}
\end{tocentry}

\begin{abstract}
It is anticipated that wide-bandgap semiconductors (WBGSs) would be useful materials for energy production and storage. A well-synthesized, yet, scarcely explored diamond-like quaternary semiconductor-Li$_2$ZnGeS$_4$ has been considered for this work. Herein, we have employed two well-known functionals GGA and mGGA within a framework of density functional theory (DFT). We have explored the electronic,  optical, mechanical, and piezo-electromechanical properties. Our results are in qualitative agreement with some of the previously reported data. The structural stabilities have been confirmed using the Born stability criteria and Molecular-dynamic (MD) simulations. Based on our findings, we claim that Li$_2$ZnGeS$_4$ is the most probable candidate for optoelectronics and piezoelectric applications.
\end{abstract}

\section{Introduction}
 Solid states laser offers multifunctional characteristics in real-time applications such as in defense systems, cutting and welding in industries, various surgeries in the medical field, fiber optic networks in information communications, etc. \cite{Byer1988a, Zhang2024b} Diamond-like quaternary semiconductors(DLQSs) are well-established materials in various optoelectronic technologies, such as in non-linear optics (NLO)\cite{Li2011}, spintronics\cite{Chambers2003}, solar cell\cite{Bedjaoui2017}, thermoelectric\cite{Sevik2010} and so on\cite{Salik2020}. Recent findings highlight the potentiality of  DLQSs in Infrared Laser Frequency (ILF) conversion due to their non-linear optical responses. The other advantage is high laser-damage thresholds (LDTs) owing to their wide bandgap (WBG)\cite{Wu2017}. In the past few years, several experimental and theoretical studies have been performed on DLQSs type materials towards various applications including laser studies\cite{Huang2019, Levcenco2011} and energy harvest\cite{Berri2022, Steinhagen2009}. However, through our rigorous literature survey, we have noticed a limited amount of research work is done to explore its functionality in the field of optoelectronic and piezoelectricity.
 Hence, we intend to perform a detailed study of WBG-DLQS material for its application in light-harvesting technology and piezoelectricity.
 \par  On the account of energy crisis, the production of energy via conventional energy sources like wind, nuclear, biomass, and fossil fuels is also increasing every day. Nevertheless, it has failed to meet the ratio of energy demand and energy supply. The necessity to tap energy resources and convert them into usable form in every feasible way is at top priority\cite{Celestine2024g}. 
 %So, in order to sustain and secure the future generations, garnering of energy using every possible method is required. 
 The development of solar panels in tapping the sunlight for harvesting energy is highly innovative and green approach. This method has drawn the attention of researchers and businesses as its sources come from the inexhaustible Sun\cite{Hao2022, Cumbajin2023}. Solar cells (SC) are constructed from semiconducting materials such as Silicon-based SC\cite{Ballif2022}, Perovskite-based SC\cite{Lalhumhima2024a}, Sulvanite-based SC\cite{Lalroliana2024} and even the present proposed DLQS materials\cite{Azam2015}. Although harvesting energy via solar panels is the most recognized and widely used alternative technique\cite{Eperon2016a}, it has some drawbacks like storing energy, occupying large space and disposal of non-degradable solar wafers after the expiry.\cite{Rahaman2018e, Ahmad2024}. Alternative methods of energy harvesting are crucial in terms of efficiency (faster conversion), long-lasting storage, size compatibility (portability), mobility, and durability. Numerous studies employing various methodologies have been conducted over the past decades\cite{Ibrahim2011b, Huangpeng2021}. Among several energy-generating systems, piezoelectricity which converts mechanical stress into electrical energy, is regarded as a prominent green energy production technique due to its non-emissive processes, adaptability and durability\cite{Hao2019d}. Mechanical sources like hydraulic, fluid, vibrations, and acoustics are always available and are not constrained by external environmental conditions. Piezoelectricity transforms energy without depending on external factors like weather, time, or sunlight\cite{Park2020c, Jeong2017c}.
 
 \par One of the criteria to verify the energy-harvesting potentiality of piezoelectric material is the finite value of piezoelectric constants. The piezoelectric constants depend on the polarizability of a material, credential to the polarization of electric dipole moment, and mainly occur in the structures having non-centrosymmetric atomic arrangements\cite{Ahmed2019}. Generating electrical energy through physical or mechanical stress without external interference manifests piezoelectric effects\cite{Bowen2014c}. The effect was first demonstrated by the Curie brothers in the 1880s\cite{Curie1880}. Since then the occurrence of this phenomenon has been found in several materials including silicon carbide (SiC)\cite{Mastropaolo2010}, perovskite-based semiconductors (PBS), zinc sulfide (ZnS) and so on\cite{Rana2022}. A wide bandgap material with higher dielectric polarization is crucial for generating high piezoelectric responses due to the presence of a longer lifetime of electron-hole pair. Most of the high-efficiency piezoelectric materials are Pb-based which are toxic in nature\cite{Yan2021}. So, synthesis and exploration of Pb-free material without compromising their efficiency is a big research challenge\cite{Luo2022a}. Piezoelectric functionality has been integrated in wide range of applications such as detector and generation of sonar waves (sensors)\cite{Huang2023}, voltage converters in loudspeakers and ejection of ink in inkjet printers (actuators)\cite{Zhou2024}, medical\cite{Labanca2008, Hoigne2006}, nano-robot\cite{Kapat2020}, MEMS\cite{Liu2025}, etc.
 \par In the present study, we utilised a diamond-like quaternary semiconductor material that follows the chemical formula of A$_2$BB'X$_4$ where the proposed combination of atoms comprises two Li-atoms in the A-position, one Zn-molecule at the B-position, another one Ge-atom at the B' site and four S-atoms are placed at the X-site which forms Li$_2$ZnGeS$_4$ (LZGS). However, the proposed DLQS material was first synthesized using X-ray powder diffraction (XRD) in 1999 by Kanno et.al\cite{Kanno2000}. In their work, they reported only the structure type and the lattice parameters (orthorhombic phase). Later, in 2019, Huang et.al\cite{Huang2019} synthesized the same system with a much higher temperature using the solid-state method. This was later validated using the powder XRD measurement and achieved results that are aligned with those of Kanno's work but differ in structure (monoclinic phase). As previously stated, the wide bandgap feature of these DLQS, including LZGS, compounds compelling researchers in the route towards laser research\cite{Zhang2020}. In light of this, our literature review indicates that the combined study of optoelectronic and piezoelectric properties has not been properly reported alongside the electromechanical coupling constants on this Li$_2$ZnGeS$_4$. Therefore,  we present the compound's structural, electrical, optical, mechanical, and piezoelectric properties using the Density Functional Theory (DFT) paradigm. 
 
\section{Computational Methods}

 The computations in this study were brought out using density functional theory\cite{Nityananda2017b}. Herein,  we implement the algorithm of QuantumATK's linear combination of atomic orbital (LCAO)\cite{Schlipf2015c}. Interactions between electron-ion were done with two exchange-correlations: generalized gradient approximation (GGA) using Perdew Burke Ernzerhof (PBE) functional\cite{Perdew1996e} and meta-generalised gradient approximation (mGGA) that operates using the exchange functional of SCAN \cite{Karasiev2022a}. Cell optimization was achieved using the adopted limited memory Broyden-Flecther-Goldfarb-Shanno (LBFGS) formalism embedded in the quasi-Newton methods, which constructs the reverse order of the Hessian matrix, and its methodology is ideal for optimizing problems with limited variables\cite{Zhao2021b}. Throughout the geometry optimization, we have relaxed all the limitations including- the cell's unit volumes, atomic positions, and space group. Consequently, to ensure the energy convergence of the cell and structure, the stress tolerance and Hellmann-Feymann force criterion were regulated to 0.00001 eV \AA $^{-3}$ and 0.01 eV \AA $^{-1}$. Several 200 convergence steps with a maximum step of 0.2 \AA \ were given along with these parameters.
 
 \par For calculating the numerical accuracy, we have chosen the Fermi-Dirac occupation method with density mesh cut-off and broadening energy at 85 Hartree and 25 meV, respectively. A basis set (medium) of PseudoDojo potential, similar to the double zeta polarised (DZP), has been assigned for each atom— Li, Zn, Ge, and S\cite{VanSetten2018c}. To regulate the iteration, the mixing variable Hamiltonian with a tolerance factor of 0.0001 was set using the Pulaymixer method. For geometry optimization, we sampled 10×10×10 k-mesh points using the Monkhorst-Pack technique\cite{Monkhorst1976d}, whereas we employed a more dense sampling of 12×12×12 with an identical convergence basis for property computations.

\section{Results and discussion}
\subsection{Structural and Electronic Properties}
\begin{figure}[hbt!]
	\centering
	\includegraphics[height=5cm]{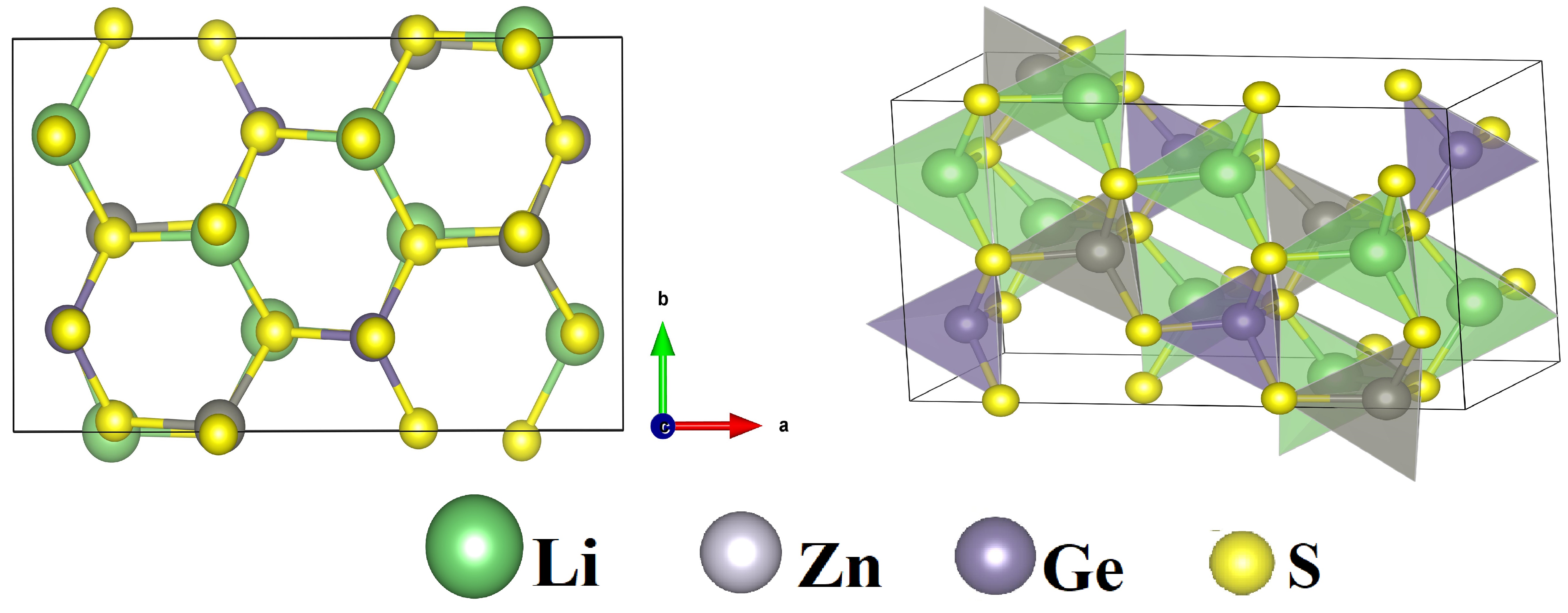}
	\caption{2D and 3D Figures of Li$_2$ZnGeS$_4$.}
	\label{Fig.1 Structure}
\end{figure}

In the present study, diamond-like quaternary compound forms an orthorhombic structure with \textit{Pna2$_1$} (33) space group. (See Fig.\ref{Fig.1 Structure}) Utilising the 3D Visualisation for Electronic and Structural Analysis (VESTA) tool\cite{Momma2008c}, the compound structure has been configured and modelled. The lattice constants, volumes, and atomic positions are fully relaxed while optimizing. After achieving the full optimization using the two employed functionals, some of the important aforementioned parameters and bandgaps are given in Table \ref{Table 1}. 

\par We observed that the optimized lattice constants of mGGA concerning GGA decrease by around 1.2\% for \textit{a}, 1.01\% for \textit{b}, and 1.42\% for \textit{c}. Also for the unit cell volumes, we obtained around 3.59\%. Along with these, we also report the bandgap percentage difference of GGA concerning mGGA which is around 11.62\%. Therefore, treating the structure using GGA achieved higher lattice constants and volumes with a smaller bandgap than that of mGGA. Our obtained results are then compared with the experimental values (See Table \ref{Table 1}). Although our findings have a lesser bandgap as compared to the other theoretical data, they are more reliable and well-suited for optoelectronic and photovoltaic-based studies.

\begin{table*}[hbt!]
	\caption{ Obtained lattice parameters (\textit{a,b,c}), volumes (\AA), and bandgaps (\textit{E$_g$})  of Li$_2$ZnGeS$_4$} 
	\begin{tabular*}{\textwidth}{@{\extracolsep{\fill}}l|l|l|l|l|l}
		\hline
		Compound(s) &Xc& \ \ \ \textit{a}\ \ \ \ \   \textit{ b}\ \ \ \ \  \textit{c} &V (\AA$^3$) & \textit{E$_g$} (eV)&  \textit{E$_g$}(Others' work) \\
		\hline			
		Li$_2$ZnGeS$_4$     & GGA  & 13.23, 7.92, 6.31 & 660.5 & 2.66&2.86\cite{Huang2019b}, 3.49*\cite{Huang2019b}\\
		Li$_2$ZnGeS$_4$   &  mGGA  & 13.07, 7.84, 6.22 &636.8&3.01&3.11\cite{Zhang2020a}, 4.07*\cite{Zhang2020a} \\
		\hline
	\end{tabular*}
'*' indicates experimental values
	\label{Table 1}
\end{table*}

\par
\begin{figure}[hbt!]
	\centering
	\includegraphics[height=7cm, width=15cm]{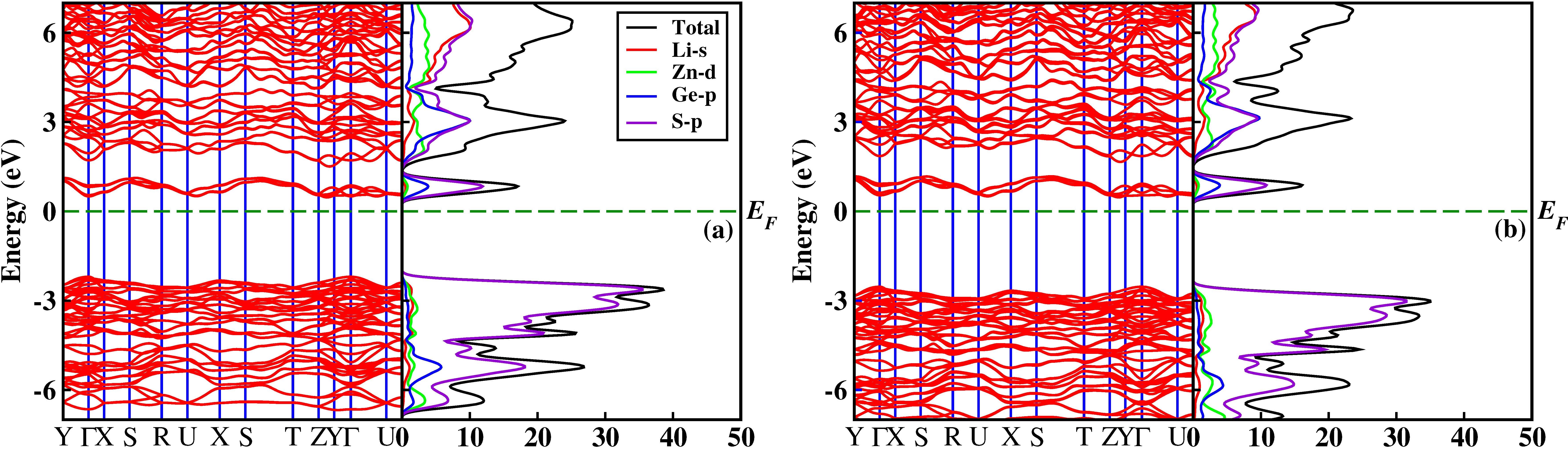}
	\caption{Band and DOS of Li$_2$ZnGeS$_4$ (a)GGA and (b)mGGA}
	\label{Fig.2 BandDOS}
\end{figure}

 The electrical characteristics of any material are one of the most important aspects to take into account while researching atomic-level relations. which uses quantum mechanics to provide a clear understanding of the interactions at the atomic level\cite{Lalhumhima2024b, Lalengmawia2025, Lalroliana2023c}. To examine the electrical characteristics of Li$_2$ZnGeS$_4$, we have employed the two exchange functionals of quantum mechanical algorithm within the framework of DFT. Depicted in Fig.\ref{Fig.2 BandDOS}, the under-investigated materials exhibit direct bandgaps confirming they are semiconductive in nature. Comparing the bandgaps under the employed functionals, we achieved the bandgap of 2.66 eV using GGA and 3.01 eV using mGGA which are then compared and agreed with the existing theoretical and experimental datas\cite{Huang2019b, Zhang2020a} (See Table \ref{Table 1}). Taking the wide bandgap into account, our review of the literature reveals that Li-based wide-bandgap semiconductors show promise as building blocks for power capacitors or batteries, which are essential for storing energy for later use\cite{Bhattacharya2018a, Tsao2018, Fan2025}. Another precedence of WBG material is that it prevents excessive charge movement that could cause an explosion or compromise the material's durability by slowing down the transfer of charge\cite{Roccaforte2018a}.

\par The band structure plots consist of conduction and valence bands ranging from -7 to +7 eV with the Fermi energy (\textit{E$_F$}) at zero-level (0 eV). As stated earlier, the compounds showed both direct bandgaps with both the highest and lowest of the band lie on the $\Gamma$-high symmetry point. Herein, from the figures, it can be seen that both band structures of Li$_2$ZnGeS$_4$ using GGA and mGGA have an n-type semiconducting nature which indicates higher concentration and greater conductivity of electrons. For the density of states (DOS) plot, the same energy range used in band structures was assigned. For better visualization, the amplitude of the DOS was set at 50 states/eV. Herein, these DOS represent the level of contribution of each atom during the electron's conductivity. From Fig.\ref{Fig.2 BandDOS}, we can observe that for both the functionals, S-atom contributes the highest in both the energy bands.

\subsection{Insights of Molecular Dynamics and Specific Heat Capacity }
\par Molecular Dynamic (MD) simulations were carried out via canonical ensemble using QuantumATK's Nose-Hoover thermostat\cite{Martyna1992d} to validate and analyze the atomic  mobilities\cite{Brooks1989b}. Since the 1960s, the usage of this simulation, using both ordered and disordered of solids, liquids, and gases have become common. Its application to solids offers the special advantage of naturally incorporating anharmonic forces and structural relaxations. The physical movements and trajectories of the system's atoms, molecules, and nanoparticles can be determined using this method. Employing the moment tensor potential predicts the physical properties of materials by assuming that particles have an interaction. The interactions between atoms are described by a variety of potentials and force fields. Having numerous uses in the domains of nanotechnology, biochemistry, and biophysics, MD can be a trustworthy tool for analysing the results of other classical models because it doesn't make any additional fundamental assumptions.
 \begin{figure*}[hbt!]
	\centering
	\includegraphics[height=5cm]{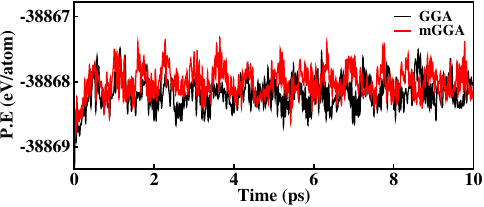}
	\caption{Molecular Dynamics Simulation of Li$_2$ZnGeS$_4$ using GGA and mGGA}
	\label{Fig.3 MD}
\end{figure*}

\par Fig.\ref{Fig.3 MD} depicts the potential energies (PE) evolution of the examined environmentally friendly perovskite upto 10-ps time steps. The nVT-based canonical ensemble\cite{Nauchitel1981} was utilised in this instance. To obtain understandable findings in the evolution of the energies, the temperature, volume, and particle count were kept constant throughout these simulations conducted at 300K. The overall energy produced by the interplay of the systems' bonding and non-bonding is represented by the PEs in Figure \ref{Fig.3 MD}. Since the simulated profiles of the PE show nearly linear variations, the compound is assumed to be thermally stable.

\begin{figure*}[hbt!]
	\centering
	\includegraphics[height=10cm]{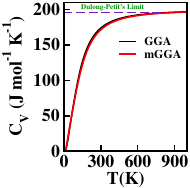}
	\caption{Specfic Heat Capacity of Li$_2$ZnGeS$_4$ using GGA and mGGA}
	\label{Fig.4 SHC}
\end{figure*}

\par A key concept of heat capacity (C$_v$) is required in comprehending a material's lattice vibrational characteristics. It refers to the amount of heat that can raise a substance's temperature by one mass unit\cite{Wolterbeek2023}. The characteristics of the C$_v$ plot curve can be loosely divided into two sections\cite{Renthlei2023h}:

\par First, when T $\ll$ $\theta_D$, it complies with the rule, which follows
\begin{equation}
C_v = \dfrac{12}{5}\pi^4 \ nR (\dfrac{T}{\theta_D})^3
\label{Eq 1}
\end{equation}

\par Secondly, when T$\gg$ $\theta_D$ the curve follows
\begin{equation}
C_v \sim 3nR
\label{Eq 2}
\end{equation}

\par where \textit{n} is the number of atoms per unit, \textit{R} represents the universal gas constant (8.314 J/K/mol) and $\theta_D$ being the Debye's temperature at 0 K.

\par Fig. \ref{Fig.4 SHC} illustrates the C$_v$ curves of Li$_2$ZnGeS$_4$ under GGA and mGGA. The rise in C$_v$ at T $\ll$ $\theta_D$ is proportional to T$^3$, indicating that it complies with Debye's T$^3$ law, a heat law particular to low temperatures. Unlike Debye's T$^3$ law, C$_v$ tends to the Dulong-Petit limit at T $\gg$ $\theta_D$, which means that it is almost constant for the system in consideration. This suggests that our calculation complies with the Dulong-Petit law of classical thermodynamics at elevated temperatures. We, thus, obtained the value of C$_v$ $\sim$ 199.536 J/K/mol for the studied DLQS system. We can therefore conclude that the compound under consideration complies with the fundamental law of thermodynamics.

\subsection{Optical Properties}
\par The optical characteristics of a substance determine how it responds to electromagnetic radiation or light waves.\cite{Lalroliana2025}. Since optical characteristics provide crucial information determining a compound's ability for photovoltaic applications, examining them is crucial to understanding its electronic configuration\cite{Krimi2024a}. We have computed a number of optical characteristics, including dielectric constants $\epsilon$($\omega$), absorption  amount $\alpha$($\omega$), and refractive indices $\eta$($\omega$) for photon energies up to +7 eV, to understand the systems' reactions to solar light. Since the under-studied compounds are asymmetric, therefore, the optical responses are obtained along xx-, yy- and zz- directions.
\par A variety of optical output outcomes, including the absorption coefficient and refractive indices, can be obtained by utilising the real $\epsilon_1$($\omega$) and imaginary $\epsilon_2$($\omega$) components of the dielectric functions.
The complex dielectric functions are expressed as follows\cite{Rahaman2018f}:
{\begin{equation}
	\epsilon(\omega) = \epsilon_{1}(\omega) + \epsilon_{2}(\omega)
	\label{Eq 3}
\end{equation}
	
\par The imaginary section $\epsilon_2$($\omega$) of the dielectric constant determines the relationship between occupied and unoccupied eigenstates associated with electronic band configurations' wave characteristics as follows:	
	
\begin{equation}
\begin{split}
\epsilon_2(\omega)= \frac{\hbar^2 e^2}{\pi m^2 \omega^2}\sum_{nn'}\int_{k}d^3k\big|\big<\vec{k}n|\vec{p}|\vec{k}n'\big>\big|\\
\times\big[1-f(\vec{k}n)\big]\delta(E_{\vec{k}n}-E_{\vec{k}n'}-\hbar\omega) 
\end{split}
\label{Eq 4}
\end{equation}	
	
\par \par where $|\vec{k}n$$\big>$ symbolize the eigen value E$_{\vec{k}n}$ , $\vec{p}$ being the operator of momentum,  and \textit{f}{($\vec{k}$n)} is the Fermi distributional function.

\par Using Kramer-Kronig's transformation\cite{Krylov2024}, the real part, $\epsilon_1$($\omega$), of the dielectric constant can be calculated from the imaginary part, $\epsilon_2$($\omega$).
\begin{equation}
\epsilon_{1}(\omega) = 1 + \dfrac{2}{\pi} \int_{0}^{\infty} \dfrac{\epsilon_{2}(\omega^{'})\omega^{'}d\omega^{'}}{\omega^{'2} - \omega^{2}}
\label{Eq 5}
\end{equation}

\par As stated earlier, the other optical properties can be obtained using the above two equations. The refractive index $\eta$($\omega$) of any material can be calculated using the following:

\begin{equation}
\eta (\omega) = \sqrt{\dfrac{[\epsilon_1^2(\omega) + \epsilon_2^2(\omega)]^\frac{1}{2}+\epsilon_1(\omega)}{2}}
\label{Eq 6}
\end{equation}

\par The absorption coefficient $\alpha$$(\omega)$ may be obtain utilizing the following equations
\begin{equation}
\alpha(\omega)= \frac{2\omega k(\omega)}{c}
\label{Eq 7}
\end{equation}

\par where ${k(\omega)}$ being the extinction coefficient
\begin{equation}
k (\omega) =\sqrt{\frac{(\epsilon_1^2 + \epsilon_2^2)^\frac{1}{2} -\epsilon_1}{2}}
\label{Eq 8}
\end{equation}

\par Making use of the functionals, we, therefore, obtained the optical properties of the studied Li$_2$ZnGeS$_4$. For all the plotted constants, the regular lines are the constants obtained using GGA and the dotted lines are for mGGA. The dielectric constants, which gauge a material's efficiency in responding to light, can be used to assess the efficacy of optoelectronic devices. Therefore, optoelectronic devices function better when the dielectric values are higher\cite{Zosiamliana2022n}.  

\par The real $\epsilon_1$($\omega$) and imaginary $\epsilon_2$($\omega$) parts of the dielectric constants are depicted in Fig.\ref{Fig.4 DC}. Firstly, we provide information about the responses of Li$_2$ZnGeS$_4$ using GGA. As can be seen in Fig.\ref{Fig.4 DC}, along different axes, the maximum value is held by the xx-axis' response whose peak reaches up to 2.22 a.u. at 2.25 eV and its least value drops at around -0.28 a.u at 3.68 eV. For yy-axis, the maximum point extends up to 1.72 a.u at 2.67 eV and the lowest value reaches 0.26 a.u at 3.21 eV, respectively. For zz-axis, the highest peak obtained is 1.86 a.u at 2.84 eV and the lowest point reaches 0.08 a.u at 3.71 eV. The negative values obtained is due to the occurrence of resonance frequency which rises due to plasmonic vibration. Secondly, utilizing the mGGA functionals, all the obtained responses are lower than that of the GGA along each different respective axes. For xx-axis, the maximum point extends up to 2.08 a.u at 3.88 eV and its minimum point drops to 0.12 a.u at 3.95 eV while the response of yy-direction observed to be the lowest among them whose highest point reaches 1.54 a.u at 3.78 eV. The response along the zz-axis reaches up to 1.68 a.u at 3.24 eV and drops to 0.21 a.u at 3.98eV, respectively.
\begin{figure*}[hbt!]
	\centering
	\includegraphics[height=10cm]{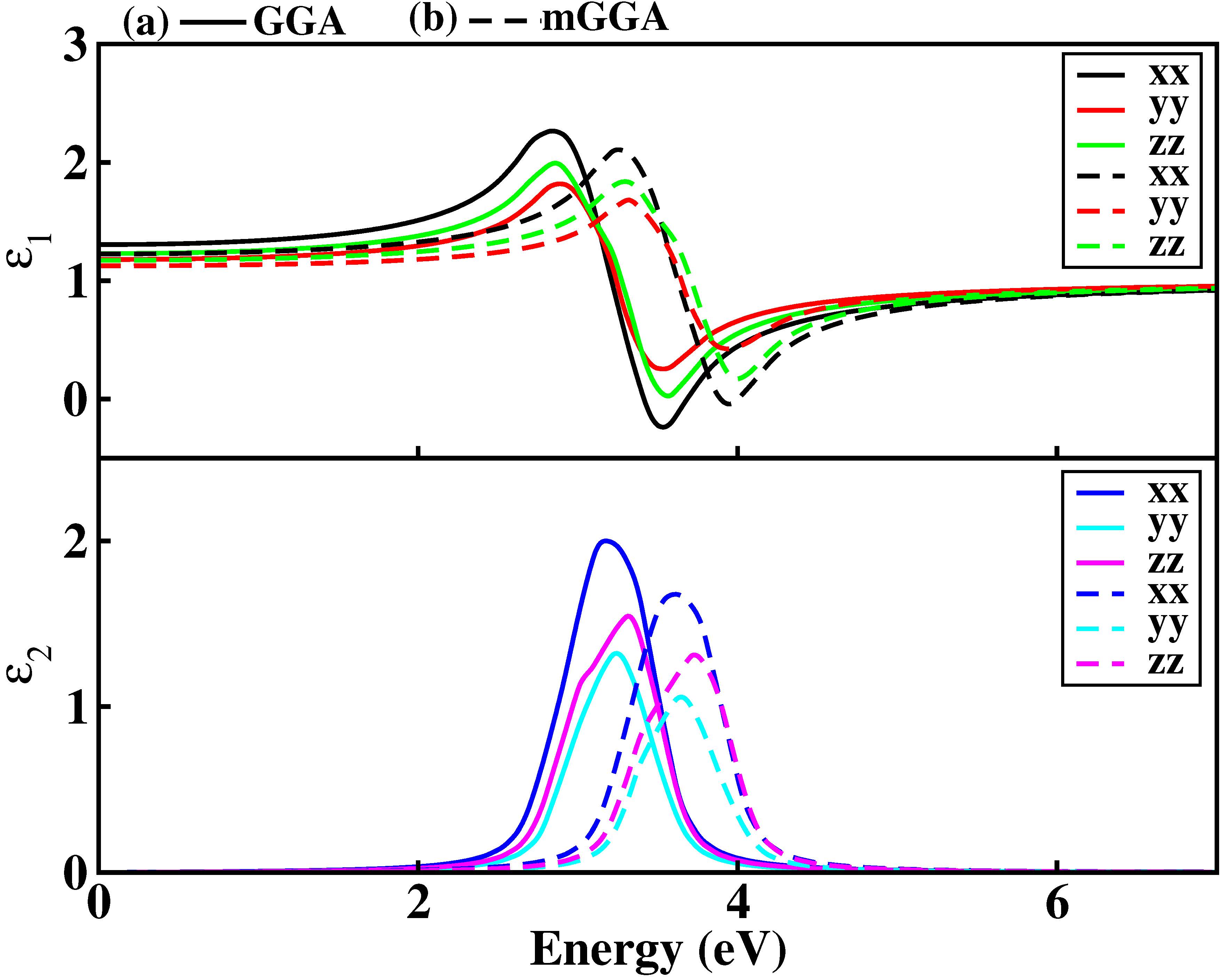}
	\caption{ Real ($\epsilon_{1}$) and imaginary ($\epsilon_{2}$) parts of the dielectric constants of the investigated Li$_2$ZnGeS$_4$ compound using GGA and mGGA}
	\label{Fig.4 DC}
\end{figure*}

\par Furthermore, the imaginary $\epsilon_2$($\omega$) parts of the dielectric constants using both the functionals have also been observed. Additionally, we have learn that in this instance, every GGA response along different axes is greater than its mGGA counterpart. For xx-direction, GGA's maximum peak extends up to 2.13 a.u at 2.29 eV while mGGA's response reached 1.85 a.u at 3.84 eV, respectively. The yy-axis for GGA reaches up to 1.26 a.u at 3.45 eV and for mGGA its response peak is observed at 1.13 a.u which falls at the level of 3.78 eV. Consequently, the peak responses along zz-axis for GGA are observed to be in 1.45 a.u at 3.16 eV and utilizing mGGA, the highest point is found to be at 1.24 a.u at 3.88 eV, respectively. In this case, the dielectric constants' imaginary components stay positive across the energy range.

\begin{figure*}[hbt!]
	\centering
	\includegraphics[height=10cm]{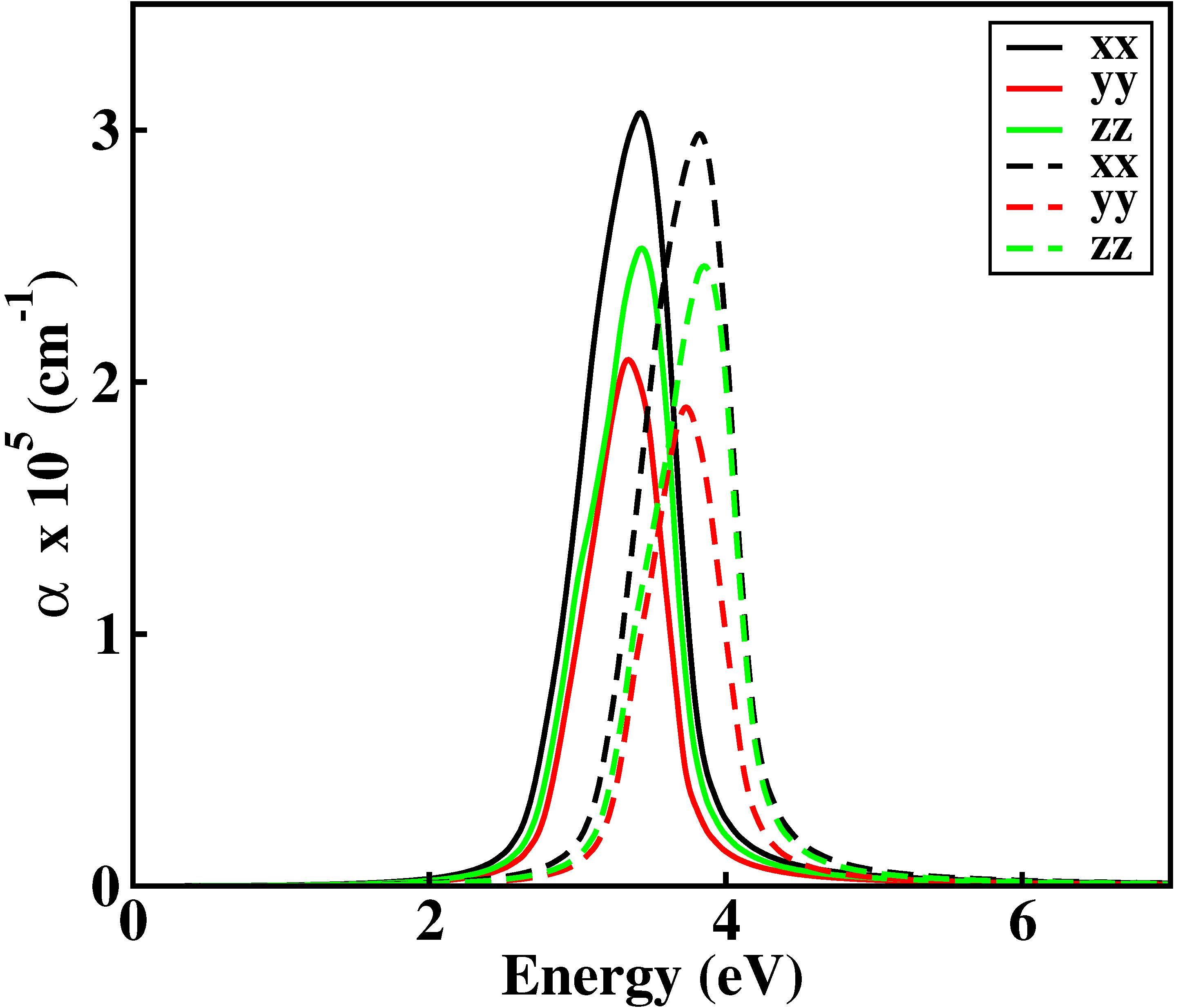}
	\caption{The absorption coefficients of Li$_2$ZnGeS$_4$ using GGA and mGGA}
	\label{Fig.5 Absorp}
\end{figure*}

\par Fig.\ref{Fig.5 Absorp} depicts the results of the absorption coefficient. We observed that all the peaks fall within the region of the vis-UV range. All the peaks obtained using GGA were found closer to the visible region and gave higher responses than the ones obtained using mGGA. Among GGA, xx-axis has the highest absorption coefficient, its absorbing capability starts from the visible region and peaks in the UV-region at around 3.05 cm$^{-1}$ at 3.62 eV and its counterpart mGGA peak is found to be 2.89 cm$^{-1}$ at  3.71 eV-. For yy-axis, the higher peak (GGA) reaches 2.05 cm$^{-1}$ at 2.87 eV while the lower peak (mGGA) extends up to 1.87 cm$^{-1}$ at 3.65 eV. Further observation reveals that for zz-axis, the response obtained from GGA reaches up to 2.51 cm$^{-1}$ at 3.32 eV and the response obtained from mGGA extends up to 2.35 cm$^{-1}$ at 3.78 eV, respectively. These findings therefore suggest that this material is a promising candidate for optoelectronics and photovoltaic applications. In addition to that, due to the presence of $\alpha$-peaks in the UV region, the materials also have the potential to be used in laser devices, ultra-violet protection devices and in numerous prospects of manufacture and healthcare applications.

\begin{figure*}[hbt!]
	\centering
	\includegraphics[height=10cm]{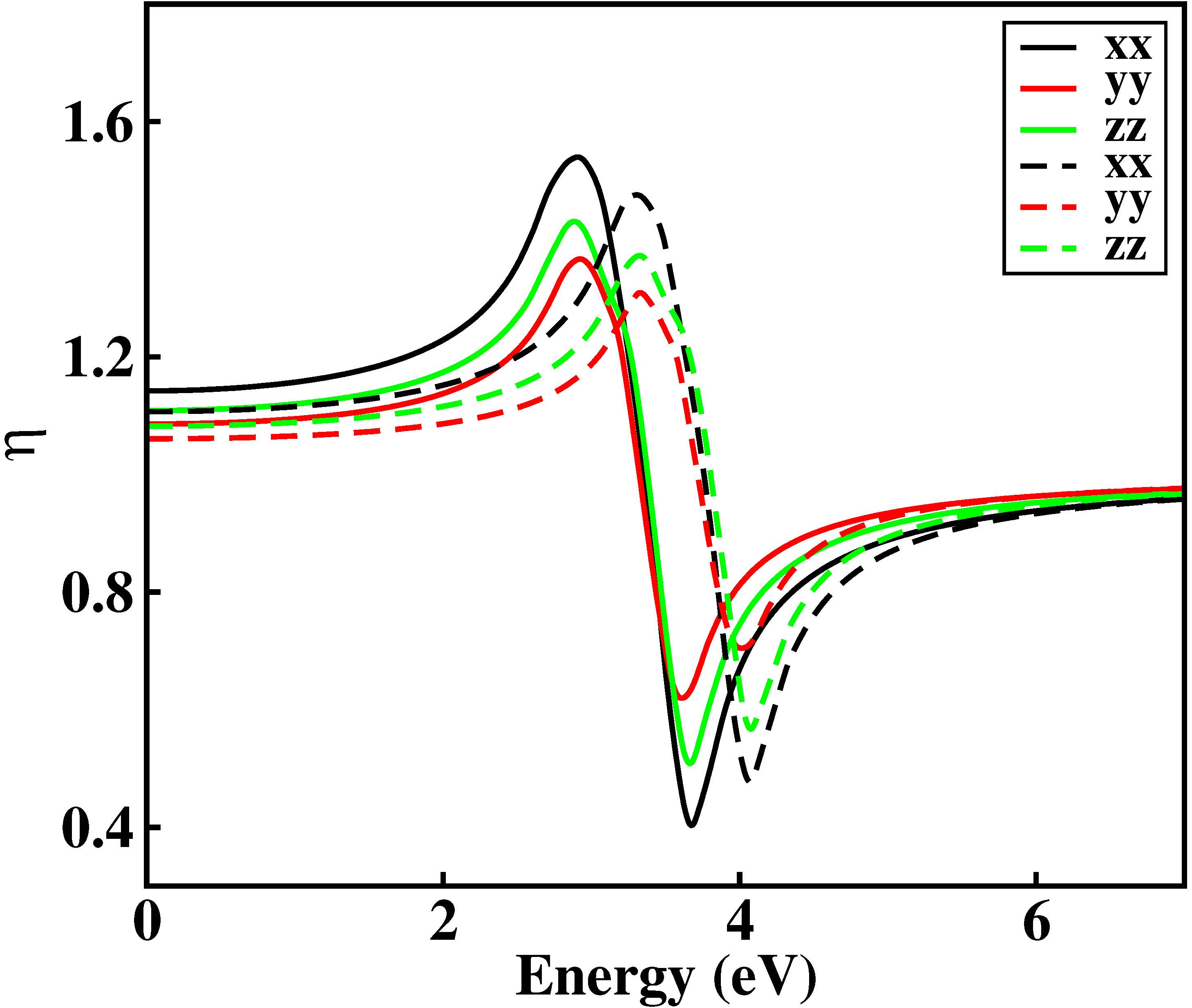}
	\caption{The calculated refractive indices of Li$_2$ZnGeS$_4$ using GGA and mGGA}
	\label{Fig.6 RI}
\end{figure*}

\par A material's ability to transmit light is described by its refractive index ($\eta$). It shows the decrease in light velocity as a material's refractive index rises\cite{Singh2002}. The interaction of atoms in the substrate and how they affect light could be the cause of this. Herein, we report the refractive indices of Li$_2$ZnGeS$_4$ using GGA and mGGA which are illustrated in Fig.\ref{Fig.6 RI}. From the figure, using GGA, we observed that xx-axis has the highest $\eta$-value of 1.57 a.u at 2.87 eV. The peak responses for yy-axis reach up to 1.28 a.u at 2.84 eV and zz-axis extends up to 1.31 a.u at 2.72 eV, respectively. Furthermore, using mGGA, it has been observed that the responses along different directions were lower than that of GGA outputs. The spectral peak for xx-axis is recorded to be 1.46 a.u at 3.85 eV. And for yy-axis, the peak reaches up to 1.21 a.u at 3.78 eV. Finally, for zz-axis, the maximum peak is obtained at 1.26 a.u along 3.77 eV, respectively. The high peaks in the derived index spectra fall inside the vis-UV region, and they closely resemble the real parts $\epsilon_1$($\omega$) of the dielectric constant. The calculated static real part of the dielectric constants and static refractive indices are listed in Table \ref{Table 2} 

\begin{table*}[hbt!]
	\caption{ Calculated static real  $\epsilon_1$$\big(0\big)$ part of the dielectric constants and static refractive index $\eta$$\big(0\big)$ along the xx-, yy-, and zz-axes of Li$_2$ZnGeS$_4$} 
	\begin{tabular*}{\textwidth}{@{\extracolsep{\fill}}l|llllll}
		\hline
	 &$\epsilon_1$$^{xx}$$\big(0\big)$ &$\epsilon_1$$^{yy}$$\big(0\big)$& $\epsilon_1$$^{zz}$$\big(0\big)$  & $\eta$$^{xx}$$\big(0\big)$ & $\eta$$^{yy}$$\big(0\big)$&$\eta$$^{zz}$$\big(0\big)$\\
		\hline	
		GGA&1.27&1.19&1.23 &1.14&1.08&1.11\\
		mGGA&1.21&1.11&1.16&1.10&1.05&1.07\\
		\hline
	\end{tabular*}
	\label{Table 2}
\end{table*}

\subsection{Elastic Constants and Mechanical Properties}

\par For practical uses in a variety of industries, a material must possess mechanical stability. A compound's mechanical stability can be ascertained by calculating its elastic constants (C$_{ij}$) using finite-strain theory\cite{Weaver1976a}. For a material to be mechanically stable under an orthorhombic symmetry, it has to fulfill the following criteria\cite{Mouhat2014}:

\begin{equation}
\begin{split}
C_{11} > 0; C_{11}C_{22} > C^2_{12} \\
C_{11}C_{22}C_{33} + 2C_{12}C_{13}C_{23} \\ 
- C_{11}C^2_{23} - C_{22}C^2_{13} \\
- C_{33}C^2_{12} > 0 \\
C_{44} > 0 ; C_{55} > 0; C_{66} > 0 \\
\end{split}
\label{Eq 9}
\end{equation}

\par The determined elastic constants given in Table \ref{Table 3} satisfy the Born stability criterion\cite{Born1940d} given in Eq.\ref{Eq 9} indicating the material is mechanically stable under GGA and mGGA. Using these elastic constants, we obtained the elastic moduli- Bulk \textit{(B)}, Young's \textit{(E)}, Shear Modulus \textit{(G)}, and so on\cite{Isotta2023}. Within the employed functionals, the calculated C$_{ij}$'s such as C$_{11}$, C$_{22}$, and C$_{33}$ are remarkably higher than that of C$_{44}$, C$_{55}$, and C$_{66}$. This implies that axial compression is more resistant to the materials being studied than shear deformation. This outcome resulted in greater bulk modulus (\textit{B}) values than shear modulus (\textit{G}) values. The calculated values of \textit{B, E, and G} are, therefore, tabulated in Table \ref{Table 3}. These moduli are important factors in determining various mechanical properties. Since these compounds have sparse the \textit{B} and \textit{G}, they can be regarded as flexible and may be suitable for thin-film manufacturing, making them suitable for use in optoelectronics.

\par We can determine the material's brittleness and ductility by using its elastic moduli. The failure mode of any material can be expressed using different measurement criteria. Utilizing the strain-dependence calculation, we derived the Poisson's ratio (\textit{v}) of the investigated material where the \textit{v} value is below(above) the critical value of 0.26, the material is said to be brittle(ductile) in nature. By computing the Pugh's ratio (\textit{k}), one can also ascertain the mode of failure of a material. In this measurement, the \textit{k} value higher(lower) than the critical value of 1.75 is regarded to be ductile(brittle). From the calculated values given in Table \ref{Table 3}, the \textit{k} and \textit{v} values read above the critical points which indicates the material, under the employed functionals, is classified to be ductile. 

\begin{table*}[hbt!]
	\small
	\caption{\ Calculated elastic constants (\textit{C$_{ij}$}), Elastic Moduli (in GPa), Pugh's ratio (\textit{k}), Poisson's ratio (\textit{v}), Gruneisen Parameter ($\zeta$), Machinability Index ($\mu_m$), Vicker's Hardness (\textit{V$_H$}), and Melting Temperature (\textit{T$_m$}) using GGA and mGGA in fully relaxed optimized structures of Li$_2$ZnGeS$_4$}
	\label{Table 3}
	\begin{tabular*}{\textwidth}{@{\extracolsep{\fill}}l|llllll}
		\hline
		& C$_{11}$ & C$_{12}$  & C$_{13}$  & C$_{22}$ & C$_{23}$ &C$_{33}$\\
		\hline
		GGA & 69.71 & 33.35 & 23.37 & 59.12 & 24.85 & 77.12\\
		mGGA & 82.41 & 40.53  & 30.29 & 70.64   &  29.35 &93.52\\
		\hline
		&  C$_{44}$  & C$_{55}$  & C$_{66}$ &\textit{B} & \textit{E}& \textit{G} \\
		\hline
		GGA &   18.15  &  15.37  & 22.40 & 40.96 &  49.37 & 19.01\\
		mGGA &   24.22   &  18.15  &  23.67   &49.47 & 58.54  & 22.47 \\
		\hline
		 &\textit{k} &\textit{v}& \textit{$\zeta$}&\textit{$\mu_m$}& \textit{V$_H$}&\textit{T$_m$}\\
		\hline			
		GGA& 2.15&0.29&0.61& 2.26 &  2.68&965.05\\
		mGGA& 2.21&0.30&0.62& 2.04& 3.01 &1040.13\\
		\hline
	\end{tabular*}
\end{table*}

\subsection{Kleinman Coefficient}
\par To understand the internal contraction stability of a compound, the Kleinmann coefficient ($\zeta$) has been utilized. This coefficient value varies between 0 to 1. The analysis shows how resistant a material is to bonding and stretching. Following the equation below, we obtained the $\zeta$ values\cite{Kleinman1962b}

\begin{equation}
\zeta = \dfrac{C_{11} + 8C_{12}}{7C_{11} + 2C_{12}}
\label{Eq 10}
\end{equation}

\par The obtained values close to one ($\zeta$ $\rightarrow$ 1) contribute more to bond bending, whereas those closer to zero ($\zeta$ $\rightarrow$ 0) influences more to bond stretching. Based on our observations, as given in Table \ref{Table 3}, our DLQS compound under the employed functionals, one could hypothesis that the examined compound's mechanical toughness lean towards bond bending.

\subsection{Machinability Index}
\par One can compute an index of machinability ($\mu_m$) to give details about a material that will be employed in industry and go through commercialization. This metric is influenced by cutting form, capacity, durability, hardness, and other machine tool-related factors. This component can be defined using the bulk modulus (\textit{B}) to shear resistance (\textit{C$_{44}$}) ratio\cite{Ahmed2023b}.

\begin{equation}
\mu_m = \dfrac{B}{C_{44}}
\label{Eq 11}
\end{equation}

\par An index ($\mu_m$) whose value higher than 1.45 can be classified as fit for the manufacturing process. Our obtained values by DLQS material surpass the standard value using both the functionals where each values are indicated in Table \ref{Table 3}. Therefore, our investigated material shows its capability in terms of device fabrication.  

\subsection{Vicker's Hardness}
\par Vicker's hardness (\textit{H$_v$}) is one of the crucial characteristics that determine the hardness of a material, particularly in the small and sparse portions. We thus obtained the values of \textit{H$_v$} using the equation\cite{Dovale-Farelo2022a}:

\begin{equation}
H_v = \dfrac{(1 - 2v)E}{6(1 + v)}
\label{Eq 12}
\end{equation}
\par where \textit{E} is the Young's modulus and \textit{v} represents the Poisson's ratio. 
\par Our calculated \textit{H$_v$} values tabulated in Table \ref{Table 3} indicate that for both the functionals employed, the toughness or hardness of this compound is found to be quite high indicating the material to be less breakable when subjecting to device fabrication.

\subsection{Melting Temperature}
\par Understanding the melting temperature ($\mu_m$) of any material is essential for its practical use.
Apart from ionic and atomic radii simulation, temperature plays a vital role in structural integrity. Using the following equation, we calculate the value of $\mu_m$

\begin{equation}
T_m = 553 + 5.911C_{11} \pm 300
\label{Eq 13}
\end{equation}

\par Our obtained $\mu_m$ values are given in Table \ref{Table 3}. We observed that under the two functionals, the investigated compound has a very high melting point indicating it can withstand high temperatures without deformation of structural stability.

\subsection{Piezoelectric Properties}
\par Piezoelectric properties have drawn a lot of interest due to their approach towards green energy harvesting capabilities\cite{Singh2023c}. The piezoelectric functions when the material is subjected to an external force which, then, produces an atomic polarization of positive and negative charges, creating an electrical energy\cite{Li2022b}. Materials with non-centrosymmetric arrangement exhibit piezoelectric response. Consequently, more polarization causes greater piezo-response\cite{King-Smith1993c}. Researchers and mineralogists have discovered lots of naturally occurring piezo-materials such as quartz, Rochelle salt and so on\cite{Saigusa2017, Valasek1922}. In contrast, artificial or man-made compounds that produce piezo-response have also been developed due to a shortage of natural piezo-materials and the need for storing and generating energy\cite{Bairagi2023}. Our proposed compound is such that it does not contain lead (Pb) element in its chemical composition, thus, considered to be viable in bio-medical applications else well\cite{Panda2009}.

\par The main objective of computing the analyzed DLQS's piezoelectric properties is to ascertain their piezoelectric tensors. Upon applying external pressure, piezoelectric materials exhibit direct electric polarization. This polarization can be reversed by generating an external electric field. Numerous applications, such as microelectromechanical devices (MEM) and systems for storing energy, can make use of these materials\cite{Bassiri-Gharb2008a}. Herein, we computed the piezoelectric tensors for this compound because there were insufficiently solid facts about the piezoelectric characteristics of the DLQS-Li$_2$ZnGeS$_4$ compound under investigation. To confirm the effectiveness and electric rate of conversion of this lead-free DLQS, we obtained electromechanical coupling constants using the computed elastic constants, dielectric constants, and the permittivity of free space in addition to these piezoelectric tensors.

\par In a bulk system without external-field influence, the sum of the spontaneous polarization P$_{eq}$ of the equilibrium structure and the strain-dependent piezoelectric polarization generated by strain P$_p$ is the total macroscopic polarization (\textit{P}), which is expressed as\cite{Bernardini1997c}

\begin{equation}
P = P_p + P_{eq}
\label{Eq 14}
\end{equation}

\par The tensors of piezoelectric can be obtained in 

\begin{equation}
\gamma_{\delta\alpha} = \dfrac{\triangle P_\delta}{\triangle \epsilon_\alpha}
\label{Eq 15}
\end{equation}

\par where the polarization \textit{P} be acquired using the method of Berry-phase approximation, and $\gamma_{\delta\alpha}$ could be computed using the QuantumATK code, which employs the finite difference method\cite{Rohrlich2009c}.

\par To determine the piezoelectric tensors using a different method, two terms must be set up: 
\begin{enumerate}
    \item The clamped-ion term e$_{i,j}$, which illustrates the electronic response to strain; 
    \item A statement that describes the impact of internal pressure on piezoelectric polarization.
\end{enumerate}

\par Consequently, the full expression for e$_{i,j}$ is obtained as 

\begin{equation}
e_{ij} = e_{ij}(0) + \dfrac{4eZ^*}{\sqrt{3}a^2} \dfrac{du}{d{\epsilon_\alpha}}
\label{Eq 16}
\end{equation}

\par where \textit{i} indicates the direction towards the applied current and \textit{j} denotes the strain direction. \textit{e} being the electronic charge, \textit{Z}* represent the Born effective charge which depends on the displacement of ion due to polarization. $\epsilon_\alpha$ denotes the macroscopic applied strain with \textit{a} as the lattice constant and \textit{u} being the interatomic distance.

\begin{table*}[hbt!]
	\small
	\caption{\ Computed piezoelectric constants of Li$_2$ZnGeS$_4$ using GGA and mGGA }
	\label{Table 4}
	\begin{tabular*}{\textwidth}{@{\extracolsep{\fill}}l|lll|l|lll}
		\hline
		GGA& x & y & z &mGGA&x& y& z\\
		\hline
		xx & \ 8.21e$^{-04}$ & 1.21e$^{-05}$ &\ 2.44e$^{-02}$ & \ \ xx&\ 3.17e$^{-03}$&\ 1.62e$^{-03}$&\ 2.23e$^{-01}$\\
		yy &  \ 2.14e$^{-04}$ & 2.29e$^{-06}$ & \ 3.04e$^{-01}$& \ \ yy&-3.21e$^{-03}$&\ 1.29e$^{-03}$&\ 2.82e$^{-01}$\\
	   
		zz&-4.53e$^{-04}$& 9.93e$^{-06}$& -5.21e$^{-01}$  & \ \ zz&\ 1.24e$^{-03}$&\ 6.32e$^{-03}$& -4.56e$^{-01}$\\		
		yz & \ 2.18e$^{-04}$ & 2.65e$^{-01}$& \ 2.31e$^{-05}$ & \ \ yz&\ 4.06e$^{-04}$&\ 2.53e$^{-01}$& -4.56e$^{-03}$\\
		xz & \ 3.39e$^{-01}$ &  2.01e$^{-04}$ &  -4.93e$^{-04}$ & \ \ xz&\ 3.25e$^{-01}$&-1.09e$^{-03}$&-4.43e$^{-03}$\\
	
		xy& -2.25e$^{-05}$ & 4.92e$^{-05}$ & -3.81e$^{-04}$ & \ \ xy&-2.54e$^{-03}$&-5.04e$^{-04}$& -4.44e$^{-04}$\\
		\hline
	\end{tabular*}
\end{table*} 

\par Table \ref{Table 4} lists the computed piezoelectric tensors for the environmentally friendly substance. Here, we used six distinct strains— xx-, yy-, zz-, yz-, xz-, and xy-strains along the x-, y-, and z- directions and obtained 18 piezoelectric tensors for the lead-free system. From our tabulated tensors, we observed that for GGA, there are 13 direct piezo-tensors and 5 inverse piezo-tensors. Whilst, we obtained  10 direct piezo-tensors and 8 inverse piezo-tensors from mGGA. From our obtained tensors, we have learn that the highest piezo-responses rose from zz-strain along the z-direction for the investigated Li$_2$ZnGeS$_4$ under GGA and mGGA. As can be seen from Table \ref{Table 4}, a response coefficient of -0.52 Cm$^{-2}$ has been achieved via GGA and another response coefficient of -0.45 Cm$^{-2}$ using mGGA, respectively. In this case, the strain is either positive or negative, as indicated by the direct and inverse effects, respectively. The average a-quartz piezo-response coefficients (0.171 Cm$^{-2}$) reported by Bechmann et al.\cite{Bechmann1958d} and the temperature-dependent values of the same a-quartz (0.07 Cm$^{-2}$) discovered by Tarumi et al.\cite{Tarumi2007c} are significantly lower than our computed piezo-response. Our investigated DLQS material reveals higher piezoelectric response than the mechanically better glass-like material (Na$_2$SiO$_3$) which was reported by Zosiamliana \textit{et al}\cite{Zosiamliana2022o}. Furthermore, our investigated compound's piezo-responses are then compared with the commonly used materials- Lithium Niobate (LiNbO$_3$) and Zinc oxide (ZnO) for piezoelectric applications like wearable electronics or MEMS sensors. We found that Chen et al.\cite{Chen2019} generated a piezo-response of 7.9 pC/N ($\sim$ 7.9e-12 Cm$^{-2}$) for LiNbO$_3$ and 3.62 pC/N ($\sim$ 3.6e-12 Cm$^{-2}$) for ZnO by Fu et al\cite{Fu2007}. In addition to these, another observation reveals that our obtained results outstand a pressurized study of CsPbBr$_3$ from 0 GPa to 15 GPa, which yielded a piezo-response from 0.14 $\mu$Ccm$^{-2}$ ($\sim$ 1.4e-7 Cm$^{-2}$) to 23.54 $\mu$Ccm$^{-2}$ ($\sim$ 2.3e-5Cm$^{-2}$) reported by Zhao et al\cite{Zhao2018}. We conclude that our calculated pielectric responses are higher than those aforementioned materials. Therefore, the DLQS material- Li$_2$ZnGeS$_4$ being studied could be used as materials for ferroelectric or piezoelectric devices.

\subsection{Electromechanical coupling constant}
\par The electromechanical coupling coefficient (\textit{k$_{ij}$}) quantifies how well a piezoelectric material converts electrical energy into mechanical energy or vice-versa. The coupling factor is computed using the piezoelectric tensor data, yielding high values. We can determine the coupling factors' efficiency using the following\cite{Roy2012d}:

\begin{equation}
\textit{k}_{ij} = \dfrac{|\textit{e}_{ij}|}{\sqrt{C_{ij}\epsilon_{fs}\epsilon_0}}
\label{Eq 17}
\end{equation}

\par where \textit{k$_{ij}$} being the electromechanical coupling coefficient. \textit{e$_{ij}$} being the obtained piezoelectric coefficient (in Cm$^{-2}$), \textit{C$_{ij}$} represents the elastic constant (in 10$^9$ Pa). $\epsilon_{fs}$ denotes the permittivity of space and $\epsilon_0$ is the static dielectric constant at the zero stress level. By using the density functional perturbed theory (DFPT), as implemented in QuantumATK, we perform a linear-response computation to determine the electromechanical coupling constants\cite{Otto1992b}.

\begin{table*}[hbt!]
	\caption{ Calculated electromechanical coupling constants
	for Li$_2$ZnGeS$_4$ where e$_{ij}$ represents the piezoelectric
	coefficients and k$_{ij}$ is the electromechanical coupling coefficient} 
	\begin{tabular*}{\textwidth}{@{\extracolsep{\fill}}l|ll}
	\hline
		& e$_{ij}$&k$_{ij}$\\
	\hline			
		GGA& -0.52&0.55\\
		mGGA& -0.45&0.44\\
	\hline
	\end{tabular*}
	\label{Table 5}
\end{table*} 

\par Making use of Eq \ref{Eq 17}, we considered the highest piezoelectric responses and other parameters required to obtain our electromechanical coupling constants from GGA and mGGA. As listed in Table \ref{Table 5}, we observed that \textit{k} = 0.55 obtained for GGA is better than \textit{k} = 0.45 calculated for mGGA.

\section{Conclusion}
\par Utilizing GGA and mGGA functionals within the DFT formalism, we thoroughly investigated the properties of the eco-friendly Li$_2$ZnGeS$_4$. The compound's ground-state stabilities have been verified based on the Born stability criteria and MD simulation. The obtained band gaps are then compared with the existing theoretical and experimental data. We report the studied DLQS material under both functionals exhibit an n-type semiconducting behaviour which facilitates quicker charge transfer. In the valence band, the S-orbital atom contributes the most. Whereas, both Li- and S- atoms are two atoms dominating in the conduction band. Properties of the optical parts are obtained using the dielectric functions. The peaks of the absorption spectra fall within the range of the vis-UV region indicating their optoelectronics potentiality. Despite the wide bandgap, the studied Li$_2$ZnGeS$_4$ with asymmetrical structure generates dielectric polarization which yielded a promising piezo-response. Using the piezoelectric tensor constants, the electromechanical coupling constants are computed. Thus, we obtained enlightening results in optical and piezoelectric tests for these infrequently researched compounds by using a distinctive technique and calculations. Consequently, we conclude that Li$_2$ZnGeS$_4$ system may be a useful energy material for technological applications.

\section{Future Scope}
\par The advantage of this DLQS material is that despite its wide bandgap and being commonly active in the field of NLO such as laser study and so on, it has the potential to be used in optoelectronic (solar cell) and piezoelectric applications. As reported in this work, the material possesses good mechanical characteristics with ductility, indicating a green flag for device fabrication. So, implying this, tuning of the bandgap via applying pressure can result in reducing the bandgap to achieve a suitable outcome towards having better results in both optoelectronics and piezoelectric applications.

\begin{acknowledgement}
Amel Laref acknowledges support from the "Research Center of the Female Scientific and Medical Colleges",  Deanship of Scientific Research, King Saud University.
\end{acknowledgement}

\section{Author Information}
\begin{enumerate}
	\item Celestine Lalengmawia: langelcelestine7@gmail.com
        \item Michael T. Nunsanga: mykiuriel@gmail.com 
         \item Saurav Suman: Saurav Suman
         \item Zosiamliana Renthlei: siama@pucollege.edu.in
	\item Lalruat Sanga: ruatsanga112@gmail.com 
        \item Hani Laltanmawii: hanivangchhia@gmail.com
        \item Lalhriat Zuala: hriata@pucollege.edu.in 
	\item Shivraj Gurung: shivraj@pucollege.edu.in
	\item Amel Laref: amel\textunderscore la06@yahoo.fr 
	\item D. P. Rai: dibyaprakashrai@gmail.com	
\end{enumerate}

\section*{Author contributions}
\textbf{Celestine Lalengmawia:} Formal analysis, Visualization, Validation, Literature review, Performed Calculation, Writing-original draft, writing-review \& editing.\\
\textbf{Michael T. Nunsanga:} Formal analysis, Visualization, Validation, writing-review \& editing. \\
\textbf{Saurav Suman:} Formal analysis, Visualization, Validation, writing-review \& editing. \\
\textbf{Zosiamliana Renthlei:} Formal analysis, Visualization, Validation, writing-review \& editing. \\
\textbf{Lalruat Sanga:} Formal analysis, Visualization, Validation, writing-review \& editing. \\
\textbf{Hani Laltanmawii:} Formal analysis, Visualization, Validation, writing-review \& editing. \\
\textbf{Lalhriat Zuala:} Formal analysis, Visualization, Validation, writing-review \& editing. \\ 
\textbf{Shivraj Gurung} Formal analysis, Visualization, Validation, writing-review \& editing. \\
\textbf{Amel Laref}: Formal analysis, Visualization, Validation, writing-review \& editing. \\
\textbf{Dibya Prakash Rai:} Project management, Supervision, Resources, software, Formal analysis, Visualization, Validation, writing-review \& editing. 

\section*{Conflicts of interest}
There are no conflicts to declare.

\section*{Data availability}
The data that support the findings of this study are available from the corresponding author upon reasonable request.

%%%%%%%%%%%%%%%%%%%%%%%%%%%%%%%%%%%%%%%%%%%%%%%%%%%%%%%%%%%%%%%%%%%%%

%%%%%%%%%%%%%%%%%%%%%%%%%%%%%%%%%%%%%%%%%%%%%%%%%%%%%%%%%%%%%%%%%%%%%

\bibliography{Li2ZnGeS4}

\providecommand{\latin}[1]{#1}
\makeatletter
\providecommand{\doi}
  {\begingroup\let\do\@makeother\dospecials
  \catcode`\{=1 \catcode`\}=2 \doi@aux}
\providecommand{\doi@aux}[1]{\endgroup\texttt{#1}}
\makeatother
\providecommand*\mcitethebibliography{\thebibliography}
\csname @ifundefined\endcsname{endmcitethebibliography}  {\let\endmcitethebibliography\endthebibliography}{}
\begin{mcitethebibliography}{96}
\providecommand*\natexlab[1]{#1}
\providecommand*\mciteSetBstSublistMode[1]{}
\providecommand*\mciteSetBstMaxWidthForm[2]{}
\providecommand*\mciteBstWouldAddEndPuncttrue
  {\def\EndOfBibitem{\unskip.}}
\providecommand*\mciteBstWouldAddEndPunctfalse
  {\let\EndOfBibitem\relax}
\providecommand*\mciteSetBstMidEndSepPunct[3]{}
\providecommand*\mciteSetBstSublistLabelBeginEnd[3]{}
\providecommand*\EndOfBibitem{}
\mciteSetBstSublistMode{f}
\mciteSetBstMaxWidthForm{subitem}{(\alph{mcitesubitemcount})}
\mciteSetBstSublistLabelBeginEnd
  {\mcitemaxwidthsubitemform\space}
  {\relax}
  {\relax}

\bibitem[Byer(1988)]{Byer1988a}
Byer,~R.~L. {Diode Laser—Pumped Solid-State Lasers}. \emph{Science} \textbf{1988}, \emph{239}, 742--747\relax
\mciteBstWouldAddEndPuncttrue
\mciteSetBstMidEndSepPunct{\mcitedefaultmidpunct}
{\mcitedefaultendpunct}{\mcitedefaultseppunct}\relax
\EndOfBibitem
\bibitem[Zhang \latin{et~al.}(2024)Zhang, Ma, Niu, Li, Wang, Wu, and Yu]{Zhang2024b}
Zhang,~Z.; Ma,~Y.; Niu,~C.; Li,~K.; Wang,~K.; Wu,~C.; Yu,~Y. {Development of all-solid-state ultraviolet lasers}. \emph{Journal of Laser Applications} \textbf{2024}, \emph{36}\relax
\mciteBstWouldAddEndPuncttrue
\mciteSetBstMidEndSepPunct{\mcitedefaultmidpunct}
{\mcitedefaultendpunct}{\mcitedefaultseppunct}\relax
\EndOfBibitem
\bibitem[Li \latin{et~al.}(2011)Li, Fan, Sun, Cheng, Li, and Zhao]{Li2011}
Li,~Y.; Fan,~W.; Sun,~H.; Cheng,~X.; Li,~P.; Zhao,~X. {Electronic, optical and lattice dynamic properties of the novel diamond-like semiconductors Li 2 CdGeS 4 and Li 2 CdSnS 4}. \emph{Journal of Physics: Condensed Matter} \textbf{2011}, \emph{23}, 225401\relax
\mciteBstWouldAddEndPuncttrue
\mciteSetBstMidEndSepPunct{\mcitedefaultmidpunct}
{\mcitedefaultendpunct}{\mcitedefaultseppunct}\relax
\EndOfBibitem
\bibitem[Liberati \latin{et~al.}(2006)Liberati, Neulinger, Chopdekar, Arenholz, Suzuki, Stacy, and Idzerda]{Chambers2003}
Liberati,~M.; Neulinger,~J.; Chopdekar,~R.; Arenholz,~E.; Suzuki,~Y.; Stacy,~A.; Idzerda,~Y. {New materials for spintronics: CuCr2Se4.} INTERMAG 2006 - IEEE International Magnetics Conference. 2006; pp 382--382\relax
\mciteBstWouldAddEndPuncttrue
\mciteSetBstMidEndSepPunct{\mcitedefaultmidpunct}
{\mcitedefaultendpunct}{\mcitedefaultseppunct}\relax
\EndOfBibitem
\bibitem[Bedjaoui \latin{et~al.}(2017)Bedjaoui, Bouhemadou, Aloumi, Khenata, Bin-Omran, Al-Douri, {Saad Saoud}, and Bensalem]{Bedjaoui2017}
Bedjaoui,~A.; Bouhemadou,~A.; Aloumi,~S.; Khenata,~R.; Bin-Omran,~S.; Al-Douri,~Y.; {Saad Saoud},~F.; Bensalem,~S. {Structural, elastic, electronic and optical properties of the novel quaternary diamond-like semiconductors Cu2MgSiS4 and Cu2MgGeS4}. \emph{Solid State Sciences} \textbf{2017}, \emph{70}, 21--35\relax
\mciteBstWouldAddEndPuncttrue
\mciteSetBstMidEndSepPunct{\mcitedefaultmidpunct}
{\mcitedefaultendpunct}{\mcitedefaultseppunct}\relax
\EndOfBibitem
\bibitem[Sevik and {\c{C}}ağın(2010)Sevik, and {\c{C}}ağın]{Sevik2010}
Sevik,~C.; {\c{C}}ağın,~T. {Ab initio study of thermoelectric transport properties of pure and doped quaternary compounds}. \emph{Physical Review B} \textbf{2010}, \emph{82}, 045202\relax
\mciteBstWouldAddEndPuncttrue
\mciteSetBstMidEndSepPunct{\mcitedefaultmidpunct}
{\mcitedefaultendpunct}{\mcitedefaultseppunct}\relax
\EndOfBibitem
\bibitem[Salik \latin{et~al.}(2020)Salik, Bouhemadou, Boudiaf, Saoud, Bin-Omran, Khenata, Al-Douri, and Reshak]{Salik2020}
Salik,~L.; Bouhemadou,~A.; Boudiaf,~K.; Saoud,~F.~S.; Bin-Omran,~S.; Khenata,~R.; Al-Douri,~Y.; Reshak,~A.~H. {Structural, elastic, electronic, magnetic, optical, and thermoelectric properties of the diamond-like quaternary semiconductor CuMn2InSe4}. \emph{Journal of Superconductivity and Novel Magnetism} \textbf{2020}, \emph{33}, 1091--1102\relax
\mciteBstWouldAddEndPuncttrue
\mciteSetBstMidEndSepPunct{\mcitedefaultmidpunct}
{\mcitedefaultendpunct}{\mcitedefaultseppunct}\relax
\EndOfBibitem
\bibitem[Wu and Pan(2017)Wu, and Pan]{Wu2017}
Wu,~K.; Pan,~S. {Li2HgMS4 (M = Si, Ge, Sn): New Quaternary Diamond-Like Semiconductors for Infrared Laser Frequency Conversion}. \emph{Crystals} \textbf{2017}, \emph{7}, 107\relax
\mciteBstWouldAddEndPuncttrue
\mciteSetBstMidEndSepPunct{\mcitedefaultmidpunct}
{\mcitedefaultendpunct}{\mcitedefaultseppunct}\relax
\EndOfBibitem
\bibitem[Huang \latin{et~al.}(2019)Huang, Wu, Cheng, Chu, Yang, and Pan]{Huang2019}
Huang,~Y.; Wu,~K.; Cheng,~J.; Chu,~Y.; Yang,~Z.; Pan,~S. {Li 2 ZnGeS 4 : a promising diamond-like infrared nonlinear optical material with high laser damage threshold and outstanding second-harmonic generation response}. \emph{Dalton Transactions} \textbf{2019}, \emph{48}, 4484--4488\relax
\mciteBstWouldAddEndPuncttrue
\mciteSetBstMidEndSepPunct{\mcitedefaultmidpunct}
{\mcitedefaultendpunct}{\mcitedefaultseppunct}\relax
\EndOfBibitem
\bibitem[Levcenco \latin{et~al.}(2011)Levcenco, Dumcenco, Huang, Arushanov, Tezlevan, Tiong, and Du]{Levcenco2011}
Levcenco,~S.; Dumcenco,~D.; Huang,~Y.; Arushanov,~E.; Tezlevan,~V.; Tiong,~K.; Du,~C. {Polarization-dependent electrolyte electroreflectance study of Cu2ZnSiS4 and Cu2ZnSiSe4 single crystals}. \emph{Journal of Alloys and Compounds} \textbf{2011}, \emph{509}, 7105--7108\relax
\mciteBstWouldAddEndPuncttrue
\mciteSetBstMidEndSepPunct{\mcitedefaultmidpunct}
{\mcitedefaultendpunct}{\mcitedefaultseppunct}\relax
\EndOfBibitem
\bibitem[Berri \latin{et~al.}(2022)Berri, Amari, Bouarissa, and Miloud]{Berri2022}
Berri,~S.; Amari,~R.; Bouarissa,~N.; Miloud,~I. {Study on quaternary diamond-like Li2CaGeO4 properties for optoelectronic applications}. \emph{Computational Condensed Matter} \textbf{2022}, \emph{30}, e00646\relax
\mciteBstWouldAddEndPuncttrue
\mciteSetBstMidEndSepPunct{\mcitedefaultmidpunct}
{\mcitedefaultendpunct}{\mcitedefaultseppunct}\relax
\EndOfBibitem
\bibitem[Steinhagen \latin{et~al.}(2009)Steinhagen, Panthani, Akhavan, Goodfellow, Koo, and Korgel]{Steinhagen2009}
Steinhagen,~C.; Panthani,~M.~G.; Akhavan,~V.; Goodfellow,~B.; Koo,~B.; Korgel,~B.~A. {Synthesis of Cu 2 ZnSnS 4 Nanocrystals for Use in Low-Cost Photovoltaics}. \emph{Journal of the American Chemical Society} \textbf{2009}, \emph{131}, 12554--12555\relax
\mciteBstWouldAddEndPuncttrue
\mciteSetBstMidEndSepPunct{\mcitedefaultmidpunct}
{\mcitedefaultendpunct}{\mcitedefaultseppunct}\relax
\EndOfBibitem
\bibitem[Celestine \latin{et~al.}(2024)Celestine, Zosiamliana, Gurung, Bhandari, Laref, Abdullaev, and Rai]{Celestine2024g}
Celestine,~L.; Zosiamliana,~R.; Gurung,~S.; Bhandari,~S.~R.; Laref,~A.; Abdullaev,~S.; Rai,~D.~P. {A Halide‐Based Perovskite CsGeX 3 (X = Cl, Br, and I) for Optoelectronic and Piezoelectric Applications}. \emph{Advanced Theory and Simulations} \textbf{2024}, \emph{7}, 2300566\relax
\mciteBstWouldAddEndPuncttrue
\mciteSetBstMidEndSepPunct{\mcitedefaultmidpunct}
{\mcitedefaultendpunct}{\mcitedefaultseppunct}\relax
\EndOfBibitem
\bibitem[Hao \latin{et~al.}(2022)Hao, Qi, Tairab, Ahmed, Azam, Luo, Pan, Zhang, and Yan]{Hao2022}
Hao,~D.; Qi,~L.; Tairab,~A.~M.; Ahmed,~A.; Azam,~A.; Luo,~D.; Pan,~Y.; Zhang,~Z.; Yan,~J. {Solar energy harvesting technologies for PV self-powered applications: A comprehensive review}. \emph{Renewable Energy} \textbf{2022}, \emph{188}, 678--697\relax
\mciteBstWouldAddEndPuncttrue
\mciteSetBstMidEndSepPunct{\mcitedefaultmidpunct}
{\mcitedefaultendpunct}{\mcitedefaultseppunct}\relax
\EndOfBibitem
\bibitem[Cumbaj{\'{i}}n \latin{et~al.}(2023)Cumbaj{\'{i}}n, S{\'{a}}nchez, N{\'{u}}{\~{n}}ez, and Gord{\'{o}}n]{Cumbajin2023}
Cumbaj{\'{i}}n,~M.; S{\'{a}}nchez,~P.; N{\'{u}}{\~{n}}ez,~M.; Gord{\'{o}}n,~C. {Energy Harvesting System with Solar Panels to Supply Low Power Electronic Devices}. \emph{IOP Conference Series: Earth and Environmental Science} \textbf{2023}, \emph{1141}, 012008\relax
\mciteBstWouldAddEndPuncttrue
\mciteSetBstMidEndSepPunct{\mcitedefaultmidpunct}
{\mcitedefaultendpunct}{\mcitedefaultseppunct}\relax
\EndOfBibitem
\bibitem[Ballif \latin{et~al.}(2022)Ballif, Haug, Boccard, Verlinden, and Hahn]{Ballif2022}
Ballif,~C.; Haug,~F.-J.; Boccard,~M.; Verlinden,~P.~J.; Hahn,~G. {Status and perspectives of crystalline silicon photovoltaics in research and industry}. \emph{Nature Reviews Materials} \textbf{2022}, \emph{7}, 597--616\relax
\mciteBstWouldAddEndPuncttrue
\mciteSetBstMidEndSepPunct{\mcitedefaultmidpunct}
{\mcitedefaultendpunct}{\mcitedefaultseppunct}\relax
\EndOfBibitem
\bibitem[Lalhumhima \latin{et~al.}(2024)Lalhumhima, Lalroliana, Lalmuanchhana, Zosiamliana, Rai, Tiwari, and Lalhriatzuala]{Lalhumhima2024a}
Lalhumhima; Lalroliana,~B.; Lalmuanchhana; Zosiamliana,~R.; Rai,~D.~P.; Tiwari,~R.~C.; Lalhriatzuala {Comprehensive investigation of structural, magnetic, electronic, optical, mechanical, and piezoelectric properties of ATiO 3 (A = Mn, Fe, Ni) compounds for sustainable energy materials}. \emph{Journal of Physics: Condensed Matter} \textbf{2024}, \emph{36}, 485901\relax
\mciteBstWouldAddEndPuncttrue
\mciteSetBstMidEndSepPunct{\mcitedefaultmidpunct}
{\mcitedefaultendpunct}{\mcitedefaultseppunct}\relax
\EndOfBibitem
\bibitem[Lalroliana \latin{et~al.}(2024)Lalroliana, Chhana, Hima, Zosiamliana, Gurung, and Zuala]{Lalroliana2024}
Lalroliana,~B.; Chhana,~L.; Hima,~L.; Zosiamliana,~R.; Gurung,~S.; Zuala,~L. {Nb-based copper sulvanites for potential green energy harvesting under induced isotropic pressure}. \emph{Materials Research Bulletin} \textbf{2024}, \emph{180}, 113054\relax
\mciteBstWouldAddEndPuncttrue
\mciteSetBstMidEndSepPunct{\mcitedefaultmidpunct}
{\mcitedefaultendpunct}{\mcitedefaultseppunct}\relax
\EndOfBibitem
\bibitem[Azam \latin{et~al.}(2015)Azam, Khan, and Goumri-Said]{Azam2015}
Azam,~S.; Khan,~S.~A.; Goumri-Said,~S. {Modified Becke–Johnson (mBJ) exchange potential investigations of the optoelectronic structure of the quaternary diamond-like semiconductors Li2CdGeS4 and Li2CdSnS4}. \emph{Materials Science in Semiconductor Processing} \textbf{2015}, \emph{39}, 606--613\relax
\mciteBstWouldAddEndPuncttrue
\mciteSetBstMidEndSepPunct{\mcitedefaultmidpunct}
{\mcitedefaultendpunct}{\mcitedefaultseppunct}\relax
\EndOfBibitem
\bibitem[Eperon \latin{et~al.}(2016)Eperon, Leijtens, Bush, Prasanna, Green, Wang, McMeekin, Volonakis, Milot, May, Palmstrom, Slotcavage, Belisle, Patel, Parrott, Sutton, Ma, Moghadam, Conings, Babayigit, Boyen, Bent, Giustino, Herz, Johnston, McGehee, and Snaith]{Eperon2016a}
Eperon,~G.~E. \latin{et~al.}  {Perovskite-perovskite tandem photovoltaics with optimized band gaps}. \emph{Science} \textbf{2016}, \emph{354}, 861--865\relax
\mciteBstWouldAddEndPuncttrue
\mciteSetBstMidEndSepPunct{\mcitedefaultmidpunct}
{\mcitedefaultendpunct}{\mcitedefaultseppunct}\relax
\EndOfBibitem
\bibitem[Rahaman and {Akther Hossain}(2018)Rahaman, and {Akther Hossain}]{Rahaman2018e}
Rahaman,~M.~Z.; {Akther Hossain},~A.~K. {Effect of metal doping on the visible light absorption, electronic structure and mechanical properties of non-toxic metal halide CsGeCl 3}. \emph{RSC Advances} \textbf{2018}, \emph{8}, 33010--33018\relax
\mciteBstWouldAddEndPuncttrue
\mciteSetBstMidEndSepPunct{\mcitedefaultmidpunct}
{\mcitedefaultendpunct}{\mcitedefaultseppunct}\relax
\EndOfBibitem
\bibitem[Ahmad \latin{et~al.}(2024)Ahmad, Zakria, Alotaibi, Huma, Hakimi, Ali, Qasimullah, Safeen, Feng, and Shah]{Ahmad2024}
Ahmad,~S.; Zakria,~M.; Alotaibi,~K.~M.; Huma,~T.; Hakimi,~N.; Ali,~R.; Qasimullah; Safeen,~A.; Feng,~J.; Shah,~S.~H. {DFT study of elastic, structural, and optical properties of K2InAgZ6 (Z = Cl, Br, I) perovskites: potential for optoelectronic applications}. \emph{Optical and Quantum Electronics} \textbf{2024}, \emph{56}, 1946\relax
\mciteBstWouldAddEndPuncttrue
\mciteSetBstMidEndSepPunct{\mcitedefaultmidpunct}
{\mcitedefaultendpunct}{\mcitedefaultseppunct}\relax
\EndOfBibitem
\bibitem[Ibrahim \latin{et~al.}(2011)Ibrahim, Ghandour, Dimitrova, Ilinca, and Perron]{Ibrahim2011b}
Ibrahim,~H.; Ghandour,~M.; Dimitrova,~M.; Ilinca,~A.; Perron,~J. {Integration of Wind Energy into Electricity Systems: Technical Challenges and Actual Solutions}. \emph{Energy Procedia} \textbf{2011}, \emph{6}, 815--824\relax
\mciteBstWouldAddEndPuncttrue
\mciteSetBstMidEndSepPunct{\mcitedefaultmidpunct}
{\mcitedefaultendpunct}{\mcitedefaultseppunct}\relax
\EndOfBibitem
\bibitem[Huangpeng \latin{et~al.}(2021)Huangpeng, Huang, and Gholinia]{Huangpeng2021}
Huangpeng,~Q.; Huang,~W.; Gholinia,~F. {Forecast of the hydropower generation under influence of climate change based on RCPs and Developed Crow Search Optimization Algorithm}. \emph{Energy Reports} \textbf{2021}, \emph{7}, 385--397\relax
\mciteBstWouldAddEndPuncttrue
\mciteSetBstMidEndSepPunct{\mcitedefaultmidpunct}
{\mcitedefaultendpunct}{\mcitedefaultseppunct}\relax
\EndOfBibitem
\bibitem[Hao \latin{et~al.}(2019)Hao, Li, Zhai, and Chen]{Hao2019d}
Hao,~J.; Li,~W.; Zhai,~J.; Chen,~H. {Progress in high-strain perovskite piezoelectric ceramics}. \emph{Materials Science and Engineering: R: Reports} \textbf{2019}, \emph{135}, 1--57\relax
\mciteBstWouldAddEndPuncttrue
\mciteSetBstMidEndSepPunct{\mcitedefaultmidpunct}
{\mcitedefaultendpunct}{\mcitedefaultseppunct}\relax
\EndOfBibitem
\bibitem[Park \latin{et~al.}(2020)Park, Ha, and Lee]{Park2020c}
Park,~H.; Ha,~C.; Lee,~J.-H. {Advances in piezoelectric halide perovskites for energy harvesting applications}. \emph{Journal of Materials Chemistry A} \textbf{2020}, \emph{8}, 24353--24367\relax
\mciteBstWouldAddEndPuncttrue
\mciteSetBstMidEndSepPunct{\mcitedefaultmidpunct}
{\mcitedefaultendpunct}{\mcitedefaultseppunct}\relax
\EndOfBibitem
\bibitem[Jeong \latin{et~al.}(2017)Jeong, Cho, Han, Park, Yang, Park, Ryu, Sohn, Chung, and Lee]{Jeong2017c}
Jeong,~C.~K.; Cho,~S.~B.; Han,~J.~H.; Park,~D.~Y.; Yang,~S.; Park,~K.-I.; Ryu,~J.; Sohn,~H.; Chung,~Y.-C.; Lee,~K.~J. {Flexible highly-effective energy harvester via crystallographic and computational control of nanointerfacial morphotropic piezoelectric thin film}. \emph{Nano Research} \textbf{2017}, \emph{10}, 437--455\relax
\mciteBstWouldAddEndPuncttrue
\mciteSetBstMidEndSepPunct{\mcitedefaultmidpunct}
{\mcitedefaultendpunct}{\mcitedefaultseppunct}\relax
\EndOfBibitem
\bibitem[Gao \latin{et~al.}(2019)Gao, Ao, and Jiang]{Ahmed2019}
Gao,~S.; Ao,~H.; Jiang,~H. {Properties and Performance of General Piezoelectric Materials on a Novel Cantilevered Energy Harvester}. \emph{IOP Conference Series: Materials Science and Engineering} \textbf{2019}, \emph{562}, 012098\relax
\mciteBstWouldAddEndPuncttrue
\mciteSetBstMidEndSepPunct{\mcitedefaultmidpunct}
{\mcitedefaultendpunct}{\mcitedefaultseppunct}\relax
\EndOfBibitem
\bibitem[Bowen \latin{et~al.}(2014)Bowen, Kim, Weaver, and Dunn]{Bowen2014c}
Bowen,~C.~R.; Kim,~H.~A.; Weaver,~P.~M.; Dunn,~S. {Piezoelectric and ferroelectric materials and structures for energy harvesting applications}. \emph{Energy Environ. Sci.} \textbf{2014}, \emph{7}, 25--44\relax
\mciteBstWouldAddEndPuncttrue
\mciteSetBstMidEndSepPunct{\mcitedefaultmidpunct}
{\mcitedefaultendpunct}{\mcitedefaultseppunct}\relax
\EndOfBibitem
\bibitem[Curie and Curie(1880)Curie, and Curie]{Curie1880}
Curie,~J.; Curie,~P. {D{\'{e}}veloppement par compression de l'{\'{e}}lectricit{\'{e}} polaire dans les cristaux h{\'{e}}mi{\`{e}}dres {\`{a}} faces inclin{\'{e}}es}. \emph{Bulletin de la Soci{\'{e}}t{\'{e}} min{\'{e}}ralogique de France} \textbf{1880}, \emph{3}, 90--93\relax
\mciteBstWouldAddEndPuncttrue
\mciteSetBstMidEndSepPunct{\mcitedefaultmidpunct}
{\mcitedefaultendpunct}{\mcitedefaultseppunct}\relax
\EndOfBibitem
\bibitem[Mastropaolo \latin{et~al.}(2010)Mastropaolo, Gual, Wood, Bunting, and Cheung]{Mastropaolo2010}
Mastropaolo,~E.; Gual,~I.; Wood,~G.; Bunting,~A.; Cheung,~R. {Piezoelectrically driven silicon carbide resonators}. \emph{Journal of Vacuum Science \& Technology B, Nanotechnology and Microelectronics: Materials, Processing, Measurement, and Phenomena} \textbf{2010}, \emph{28}, C6N18--C6N23\relax
\mciteBstWouldAddEndPuncttrue
\mciteSetBstMidEndSepPunct{\mcitedefaultmidpunct}
{\mcitedefaultendpunct}{\mcitedefaultseppunct}\relax
\EndOfBibitem
\bibitem[Rana \latin{et~al.}(2022)Rana, Khan, Zhu, Fattah, Kokilathasan, Rassel, Bernard, Ababou-Girard, Turban, Xu, Wang, and Ban]{Rana2022}
Rana,~M.~M.; Khan,~A.~A.; Zhu,~W.; Fattah,~M. F.~A.; Kokilathasan,~S.; Rassel,~S.; Bernard,~R.; Ababou-Girard,~S.; Turban,~P.; Xu,~S.; Wang,~C.; Ban,~D. {Enhanced piezoelectricity in lead-free halide perovskite nanocomposite for self-powered wireless electronics}. \emph{Nano Energy} \textbf{2022}, \emph{101}, 107631\relax
\mciteBstWouldAddEndPuncttrue
\mciteSetBstMidEndSepPunct{\mcitedefaultmidpunct}
{\mcitedefaultendpunct}{\mcitedefaultseppunct}\relax
\EndOfBibitem
\bibitem[Yan \latin{et~al.}(2021)Yan, Li, Jin, Du, Zhang, Zhang, and Hao]{Yan2021}
Yan,~Y.; Li,~Z.; Jin,~L.; Du,~H.; Zhang,~M.; Zhang,~D.; Hao,~Y. Extremely High Piezoelectric Properties in Pb-Based Ceramics through Integrating Phase Boundary and Defect Engineering. \emph{ACS Applied Materials \& Interfaces} \textbf{2021}, \emph{13}, 38517--38525, PMID: 34353025\relax
\mciteBstWouldAddEndPuncttrue
\mciteSetBstMidEndSepPunct{\mcitedefaultmidpunct}
{\mcitedefaultendpunct}{\mcitedefaultseppunct}\relax
\EndOfBibitem
\bibitem[Luo \latin{et~al.}(2022)Luo, Li, Ma, Han, Zhang, and Song]{Luo2022a}
Luo,~H.; Li,~P.; Ma,~J.; Han,~L.; Zhang,~Y.; Song,~Y. {Sustainable Pb Management in Perovskite Solar Cells toward Eco‐Friendly Development}. \emph{Advanced Energy Materials} \textbf{2022}, \emph{12}, 2201242\relax
\mciteBstWouldAddEndPuncttrue
\mciteSetBstMidEndSepPunct{\mcitedefaultmidpunct}
{\mcitedefaultendpunct}{\mcitedefaultseppunct}\relax
\EndOfBibitem
\bibitem[Huang \latin{et~al.}(2023)Huang, Gao, Hu, Shen, Jin, and Cho]{Huang2023}
Huang,~S.; Gao,~Y.; Hu,~Y.; Shen,~F.; Jin,~Z.; Cho,~Y. {Recent development of piezoelectric biosensors for physiological signal detection and machine learning assisted cardiovascular disease diagnosis}. \emph{RSC Advances} \textbf{2023}, \emph{13}, 29174--29194\relax
\mciteBstWouldAddEndPuncttrue
\mciteSetBstMidEndSepPunct{\mcitedefaultmidpunct}
{\mcitedefaultendpunct}{\mcitedefaultseppunct}\relax
\EndOfBibitem
\bibitem[Zhou \latin{et~al.}(2024)Zhou, Wu, Wang, Wang, Zhu, Sun, Huang, Wang, Huang, and Lu]{Zhou2024}
Zhou,~X.; Wu,~S.; Wang,~X.; Wang,~Z.; Zhu,~Q.; Sun,~J.; Huang,~P.; Wang,~X.; Huang,~W.; Lu,~Q. {Review on piezoelectric actuators: materials, classifications, applications, and recent trends}. \emph{Frontiers of Mechanical Engineering} \textbf{2024}, \emph{19}, 6\relax
\mciteBstWouldAddEndPuncttrue
\mciteSetBstMidEndSepPunct{\mcitedefaultmidpunct}
{\mcitedefaultendpunct}{\mcitedefaultseppunct}\relax
\EndOfBibitem
\bibitem[Labanca \latin{et~al.}(2008)Labanca, Azzola, Vinci, and Rodella]{Labanca2008}
Labanca,~M.; Azzola,~F.; Vinci,~R.; Rodella,~L.~F. {Piezoelectric surgery: Twenty years of use}. \emph{British Journal of Oral and Maxillofacial Surgery} \textbf{2008}, \emph{46}, 265--269\relax
\mciteBstWouldAddEndPuncttrue
\mciteSetBstMidEndSepPunct{\mcitedefaultmidpunct}
{\mcitedefaultendpunct}{\mcitedefaultseppunct}\relax
\EndOfBibitem
\bibitem[Hoigne \latin{et~al.}(2006)Hoigne, St{\"{u}}binger, Kaenel, Shamdasani, and Hasenboehler]{Hoigne2006}
Hoigne,~D.~J.; St{\"{u}}binger,~S.; Kaenel,~O.~V.; Shamdasani,~S.; Hasenboehler,~P. {Piezoelectric osteotomy in hand surgery: first experiences with a new technique}. \emph{BMC Musculoskeletal Disorders} \textbf{2006}, \emph{7}, 36\relax
\mciteBstWouldAddEndPuncttrue
\mciteSetBstMidEndSepPunct{\mcitedefaultmidpunct}
{\mcitedefaultendpunct}{\mcitedefaultseppunct}\relax
\EndOfBibitem
\bibitem[Kapat \latin{et~al.}(2020)Kapat, Shubhra, Zhou, and Leeuwenburgh]{Kapat2020}
Kapat,~K.; Shubhra,~Q. T.~H.; Zhou,~M.; Leeuwenburgh,~S. Piezoelectric Nano-Biomaterials for Biomedicine and Tissue Regeneration. \emph{Advanced Functional Materials} \textbf{2020}, \emph{30}, 1909045\relax
\mciteBstWouldAddEndPuncttrue
\mciteSetBstMidEndSepPunct{\mcitedefaultmidpunct}
{\mcitedefaultendpunct}{\mcitedefaultseppunct}\relax
\EndOfBibitem
\bibitem[Liu \latin{et~al.}(2025)Liu, Tan, Zhou, Ma, Wang, Tran, Lu, Chen, Wang, and Zhang]{Liu2025}
Liu,~J.; Tan,~H.; Zhou,~X.; Ma,~W.; Wang,~C.; Tran,~N.-M.-A.; Lu,~W.; Chen,~F.; Wang,~J.; Zhang,~H. Piezoelectric thin films and their applications in MEMS: A review. \emph{Journal of Applied Physics} \textbf{2025}, \emph{137}, 020702\relax
\mciteBstWouldAddEndPuncttrue
\mciteSetBstMidEndSepPunct{\mcitedefaultmidpunct}
{\mcitedefaultendpunct}{\mcitedefaultseppunct}\relax
\EndOfBibitem
\bibitem[Kanno \latin{et~al.}(2000)Kanno, Hata, Kawamoto, and Irie]{Kanno2000}
Kanno,~R.; Hata,~T.; Kawamoto,~Y.; Irie,~M. {Synthesis of a new lithium ionic conductor, thio-LISICON–lithium germanium sulfide system}. \emph{Solid State Ionics} \textbf{2000}, \emph{130}, 97--104\relax
\mciteBstWouldAddEndPuncttrue
\mciteSetBstMidEndSepPunct{\mcitedefaultmidpunct}
{\mcitedefaultendpunct}{\mcitedefaultseppunct}\relax
\EndOfBibitem
\bibitem[Zhang \latin{et~al.}(2020)Zhang, Clark, Brant, Rosmus, Grima, Lekse, Jang, and Aitken]{Zhang2020}
Zhang,~J.-H.; Clark,~D.~J.; Brant,~J.~A.; Rosmus,~K.~A.; Grima,~P.; Lekse,~J.~W.; Jang,~J.~I.; Aitken,~J.~A. {$\alpha$-Li 2 ZnGeS 4 : A Wide-Bandgap Diamond-like Semiconductor with Excellent Balance between Laser-Induced Damage Threshold and Second Harmonic Generation Response}. \emph{Chemistry of Materials} \textbf{2020}, \emph{32}, 8947--8955\relax
\mciteBstWouldAddEndPuncttrue
\mciteSetBstMidEndSepPunct{\mcitedefaultmidpunct}
{\mcitedefaultendpunct}{\mcitedefaultseppunct}\relax
\EndOfBibitem
\bibitem[Nityananda \latin{et~al.}(2017)Nityananda, Hohenberg, and Kohn]{Nityananda2017b}
Nityananda,~R.; Hohenberg,~P.; Kohn,~W. {Inhomogeneous electron gas}. \emph{Resonance} \textbf{2017}, \emph{22}, 809--811\relax
\mciteBstWouldAddEndPuncttrue
\mciteSetBstMidEndSepPunct{\mcitedefaultmidpunct}
{\mcitedefaultendpunct}{\mcitedefaultseppunct}\relax
\EndOfBibitem
\bibitem[Schlipf and Gygi(2015)Schlipf, and Gygi]{Schlipf2015c}
Schlipf,~M.; Gygi,~F. {Optimization algorithm for the generation of ONCV pseudopotentials}. \emph{Computer Physics Communications} \textbf{2015}, \emph{196}, 36--44\relax
\mciteBstWouldAddEndPuncttrue
\mciteSetBstMidEndSepPunct{\mcitedefaultmidpunct}
{\mcitedefaultendpunct}{\mcitedefaultseppunct}\relax
\EndOfBibitem
\bibitem[Perdew \latin{et~al.}(1996)Perdew, Burke, and Ernzerhof]{Perdew1996e}
Perdew,~J.~P.; Burke,~K.; Ernzerhof,~M. {Generalized Gradient Approximation Made Simple}. \emph{Physical Review Letters} \textbf{1996}, \emph{77}, 3865--3868\relax
\mciteBstWouldAddEndPuncttrue
\mciteSetBstMidEndSepPunct{\mcitedefaultmidpunct}
{\mcitedefaultendpunct}{\mcitedefaultseppunct}\relax
\EndOfBibitem
\bibitem[Karasiev \latin{et~al.}(2022)Karasiev, Mihaylov, and Hu]{Karasiev2022a}
Karasiev,~V.~V.; Mihaylov,~D.~I.; Hu,~S.~X. {Meta-GGA exchange-correlation free energy density functional to increase the accuracy of warm dense matter simulations}. \emph{Physical Review B} \textbf{2022}, \emph{105}, L081109\relax
\mciteBstWouldAddEndPuncttrue
\mciteSetBstMidEndSepPunct{\mcitedefaultmidpunct}
{\mcitedefaultendpunct}{\mcitedefaultseppunct}\relax
\EndOfBibitem
\bibitem[Zhao(2021)]{Zhao2021b}
Zhao,~W. {A Broyden–Fletcher–Goldfarb–Shanno algorithm for reliability-based design optimization}. \emph{Applied Mathematical Modelling} \textbf{2021}, \emph{92}, 447--465\relax
\mciteBstWouldAddEndPuncttrue
\mciteSetBstMidEndSepPunct{\mcitedefaultmidpunct}
{\mcitedefaultendpunct}{\mcitedefaultseppunct}\relax
\EndOfBibitem
\bibitem[van Setten \latin{et~al.}(2018)van Setten, Giantomassi, Bousquet, Verstraete, Hamann, Gonze, and Rignanese]{VanSetten2018c}
van Setten,~M.~J.; Giantomassi,~M.; Bousquet,~E.; Verstraete,~M.~J.; Hamann,~D.~R.; Gonze,~X.; Rignanese,~G.~M. {The PSEUDODOJO: Training and grading a 85 element optimized norm-conserving pseudopotential table}. \emph{Computer Physics Communications} \textbf{2018}, \emph{226}, 39--54\relax
\mciteBstWouldAddEndPuncttrue
\mciteSetBstMidEndSepPunct{\mcitedefaultmidpunct}
{\mcitedefaultendpunct}{\mcitedefaultseppunct}\relax
\EndOfBibitem
\bibitem[Monkhorst and Pack(1976)Monkhorst, and Pack]{Monkhorst1976d}
Monkhorst,~H.~J.; Pack,~J.~D. {Special points for Brillouin-zone integrations}. \emph{Physical Review B} \textbf{1976}, \emph{13}, 5188--5192\relax
\mciteBstWouldAddEndPuncttrue
\mciteSetBstMidEndSepPunct{\mcitedefaultmidpunct}
{\mcitedefaultendpunct}{\mcitedefaultseppunct}\relax
\EndOfBibitem
\bibitem[Momma and Izumi(2008)Momma, and Izumi]{Momma2008c}
Momma,~K.; Izumi,~F. {VESTA : a three-dimensional visualization system for electronic and structural analysis}. \emph{Journal of Applied Crystallography} \textbf{2008}, \emph{41}, 653--658\relax
\mciteBstWouldAddEndPuncttrue
\mciteSetBstMidEndSepPunct{\mcitedefaultmidpunct}
{\mcitedefaultendpunct}{\mcitedefaultseppunct}\relax
\EndOfBibitem
\bibitem[Huang \latin{et~al.}(2019)Huang, Wu, Cheng, Chu, Yang, and Pan]{Huang2019b}
Huang,~Y.; Wu,~K.; Cheng,~J.; Chu,~Y.; Yang,~Z.; Pan,~S. {Li 2 ZnGeS 4 : a promising diamond-like infrared nonlinear optical material with high laser damage threshold and outstanding second-harmonic generation response}. \emph{Dalton Transactions} \textbf{2019}, \emph{48}, 4484--4488\relax
\mciteBstWouldAddEndPuncttrue
\mciteSetBstMidEndSepPunct{\mcitedefaultmidpunct}
{\mcitedefaultendpunct}{\mcitedefaultseppunct}\relax
\EndOfBibitem
\bibitem[Zhang \latin{et~al.}(2020)Zhang, Clark, Brant, Rosmus, Grima, Lekse, Jang, and Aitken]{Zhang2020a}
Zhang,~J.-H.; Clark,~D.~J.; Brant,~J.~A.; Rosmus,~K.~A.; Grima,~P.; Lekse,~J.~W.; Jang,~J.~I.; Aitken,~J.~A. {$\alpha$-Li 2 ZnGeS 4 : A Wide-Bandgap Diamond-like Semiconductor with Excellent Balance between Laser-Induced Damage Threshold and Second Harmonic Generation Response}. \emph{Chemistry of Materials} \textbf{2020}, \emph{32}, 8947--8955\relax
\mciteBstWouldAddEndPuncttrue
\mciteSetBstMidEndSepPunct{\mcitedefaultmidpunct}
{\mcitedefaultendpunct}{\mcitedefaultseppunct}\relax
\EndOfBibitem
\bibitem[Lalhumhima \latin{et~al.}(2024)Lalhumhima, Lalroliana, Lalmuanchhana, Zosiamliana, Rai, Tiwari, and Lalhriatzuala]{Lalhumhima2024b}
Lalhumhima; Lalroliana,~B.; Lalmuanchhana; Zosiamliana,~R.; Rai,~D.~P.; Tiwari,~R.~C.; Lalhriatzuala {Comprehensive investigation of structural, magnetic, electronic, optical, mechanical, and piezoelectric properties of ATiO 3 (A = Mn, Fe, Ni) compounds for sustainable energy materials}. \emph{Journal of Physics: Condensed Matter} \textbf{2024}, \emph{36}, 485901\relax
\mciteBstWouldAddEndPuncttrue
\mciteSetBstMidEndSepPunct{\mcitedefaultmidpunct}
{\mcitedefaultendpunct}{\mcitedefaultseppunct}\relax
\EndOfBibitem
\bibitem[Lalengmawia \latin{et~al.}(2025)Lalengmawia, Renthlei, Gurung, Zuala, Pachuau, Singh, Vanchhawng, Gopi, Yvaz, and Rai]{Lalengmawia2025}
Lalengmawia,~C.; Renthlei,~Z.; Gurung,~S.; Zuala,~L.; Pachuau,~L.; Singh,~N.~S.; Vanchhawng,~L.; Gopi,~K.; Yvaz,~A.; Rai,~D.~P. {A comprehensive study of electronic, optical, mechanical and piezoelectric properties of Li-based tin-halide perovskites using GGA, meta-GGA and HSE06 methods}. \emph{New Journal of Chemistry} \textbf{2025}, \emph{49}, 3578--3589\relax
\mciteBstWouldAddEndPuncttrue
\mciteSetBstMidEndSepPunct{\mcitedefaultmidpunct}
{\mcitedefaultendpunct}{\mcitedefaultseppunct}\relax
\EndOfBibitem
\bibitem[Lalroliana \latin{et~al.}(2023)Lalroliana, Lalmuanchhana, Lalhumhima, Lalrinkima, Gurung, {Rangeela Devi}, {Surajkumar Singh}, Zodinmawia, {Prakash Rai}, and Lalhriatzuala]{Lalroliana2023c}
Lalroliana,~B.; Lalmuanchhana; Lalhumhima; Lalrinkima; Gurung,~S.; {Rangeela Devi},~Y.; {Surajkumar Singh},~N.; Zodinmawia; {Prakash Rai},~D.; Lalhriatzuala {Ab initio prediction of half-metallicity, stability and reconstruction in Cu3TaTe4 (1 0 0) surface}. \emph{Computational Materials Science} \textbf{2023}, \emph{230}, 112476\relax
\mciteBstWouldAddEndPuncttrue
\mciteSetBstMidEndSepPunct{\mcitedefaultmidpunct}
{\mcitedefaultendpunct}{\mcitedefaultseppunct}\relax
\EndOfBibitem
\bibitem[Bhattacharya(2019)]{Bhattacharya2018a}
Bhattacharya,~S. \emph{Wide Bandgap Semiconductor Power Devices}; Elsevier, 2019; pp 249--300\relax
\mciteBstWouldAddEndPuncttrue
\mciteSetBstMidEndSepPunct{\mcitedefaultmidpunct}
{\mcitedefaultendpunct}{\mcitedefaultseppunct}\relax
\EndOfBibitem
\bibitem[Tsao \latin{et~al.}(2018)Tsao, Chowdhury, Hollis, Jena, Johnson, Jones, Kaplar, Rajan, {Van de Walle}, Bellotti, Chua, Collazo, Coltrin, Cooper, Evans, Graham, Grotjohn, Heller, Higashiwaki, Islam, Juodawlkis, Khan, Koehler, Leach, Mishra, Nemanich, Pilawa‐Podgurski, Shealy, Sitar, Tadjer, Witulski, Wraback, and Simmons]{Tsao2018}
Tsao,~J.~Y. \latin{et~al.}  {Ultrawide‐Bandgap Semiconductors: Research Opportunities and Challenges}. \emph{Advanced Electronic Materials} \textbf{2018}, \emph{4}, 1600501\relax
\mciteBstWouldAddEndPuncttrue
\mciteSetBstMidEndSepPunct{\mcitedefaultmidpunct}
{\mcitedefaultendpunct}{\mcitedefaultseppunct}\relax
\EndOfBibitem
\bibitem[Fan \latin{et~al.}(2025)Fan, Qu, Xu, Yang, Yang, Lin, Tang, and Liu]{Fan2025}
Fan,~Q.; Qu,~D.; Xu,~C.; Yang,~H.; Yang,~S.; Lin,~D.; Tang,~H.; Liu,~D. {A lithium-ion battery system with high power and wide temperature range targeting the internet of things applications}. \emph{Journal of Power Sources} \textbf{2025}, \emph{630}, 236070\relax
\mciteBstWouldAddEndPuncttrue
\mciteSetBstMidEndSepPunct{\mcitedefaultmidpunct}
{\mcitedefaultendpunct}{\mcitedefaultseppunct}\relax
\EndOfBibitem
\bibitem[Roccaforte \latin{et~al.}(2018)Roccaforte, Fiorenza, Greco, {Lo Nigro}, Giannazzo, Iucolano, and Saggio]{Roccaforte2018a}
Roccaforte,~F.; Fiorenza,~P.; Greco,~G.; {Lo Nigro},~R.; Giannazzo,~F.; Iucolano,~F.; Saggio,~M. {Emerging trends in wide band gap semiconductors (SiC and GaN) technology for power devices}. \emph{Microelectronic Engineering} \textbf{2018}, \emph{187-188}, 66--77\relax
\mciteBstWouldAddEndPuncttrue
\mciteSetBstMidEndSepPunct{\mcitedefaultmidpunct}
{\mcitedefaultendpunct}{\mcitedefaultseppunct}\relax
\EndOfBibitem
\bibitem[Martyna \latin{et~al.}(1992)Martyna, Klein, and Tuckerman]{Martyna1992d}
Martyna,~G.~J.; Klein,~M.~L.; Tuckerman,~M. {Nos{\'{e}}–Hoover chains: The canonical ensemble via continuous dynamics}. \emph{The Journal of Chemical Physics} \textbf{1992}, \emph{97}, 2635--2643\relax
\mciteBstWouldAddEndPuncttrue
\mciteSetBstMidEndSepPunct{\mcitedefaultmidpunct}
{\mcitedefaultendpunct}{\mcitedefaultseppunct}\relax
\EndOfBibitem
\bibitem[Brooks(1989)]{Brooks1989b}
Brooks,~C.~L. {Computer simulation of liquids}. \emph{Journal of Solution Chemistry} \textbf{1989}, \emph{18}, 99--99\relax
\mciteBstWouldAddEndPuncttrue
\mciteSetBstMidEndSepPunct{\mcitedefaultmidpunct}
{\mcitedefaultendpunct}{\mcitedefaultseppunct}\relax
\EndOfBibitem
\bibitem[Nauchitel'(1981)]{Nauchitel1981}
Nauchitel',~V. {Energy distribution function for the NVT canonical ensemble}. \emph{Molecular Physics} \textbf{1981}, \emph{42}, 1259--1265\relax
\mciteBstWouldAddEndPuncttrue
\mciteSetBstMidEndSepPunct{\mcitedefaultmidpunct}
{\mcitedefaultendpunct}{\mcitedefaultseppunct}\relax
\EndOfBibitem
\bibitem[Wolterbeek and Hangx(2023)Wolterbeek, and Hangx]{Wolterbeek2023}
Wolterbeek,~T.; Hangx,~S. {The thermal properties of set Portland cements – a literature review in the context of CO2 injection well integrity}. \emph{International Journal of Greenhouse Gas Control} \textbf{2023}, \emph{126}, 103909\relax
\mciteBstWouldAddEndPuncttrue
\mciteSetBstMidEndSepPunct{\mcitedefaultmidpunct}
{\mcitedefaultendpunct}{\mcitedefaultseppunct}\relax
\EndOfBibitem
\bibitem[Renthlei \latin{et~al.}(2023)Renthlei, Celestine, Kima, Zuala, Mawia, Chettri, Singh, Abdullaev, Ezzeldien, and Rai]{Renthlei2023h}
Renthlei,~Z.; Celestine,~L.; Kima,~L.; Zuala,~L.; Mawia,~Z.; Chettri,~B.; Singh,~Y.~T.; Abdullaev,~S.; Ezzeldien,~M.; Rai,~D.~P. {Theoretical Investigation of Lead Perovskite PbXO 3 (X = Ti, Zr, and Hf) for Potential Thermoelectric Applications: Hybrid-DFT Approach}. \emph{Energy \& Fuels} \textbf{2023}, \emph{37}, 19831--19844\relax
\mciteBstWouldAddEndPuncttrue
\mciteSetBstMidEndSepPunct{\mcitedefaultmidpunct}
{\mcitedefaultendpunct}{\mcitedefaultseppunct}\relax
\EndOfBibitem
\bibitem[Lalroliana \latin{et~al.}(2025)Lalroliana, Lalmuanchhana, Lalhumhima, Celestine, Rai, Pachuau, Singh, Gurung, and Lalhriatzuala]{Lalroliana2025}
Lalroliana,~B.; Lalmuanchhana; Lalhumhima; Celestine,~L.; Rai,~D.~P.; Pachuau,~L.; Singh,~N.~S.; Gurung,~S.; Lalhriatzuala {Enhanced optical and thermoelectric properties of Cu 3 Nb 1− X V X S 4 through chemical substitution: a DFT approach}. \emph{New Journal of Chemistry} \textbf{2025}, \emph{49}, 1763--1772\relax
\mciteBstWouldAddEndPuncttrue
\mciteSetBstMidEndSepPunct{\mcitedefaultmidpunct}
{\mcitedefaultendpunct}{\mcitedefaultseppunct}\relax
\EndOfBibitem
\bibitem[Krimi \latin{et~al.}(2024)Krimi, Hajlaoui, Abdelbaky, Garcia-Granda, and {Ben Rhaiem}]{Krimi2024a}
Krimi,~M.; Hajlaoui,~F.; Abdelbaky,~M. S.~M.; Garcia-Granda,~S.; {Ben Rhaiem},~A. {Investigation of optical, dielectric, and conduction mechanism in lead-free perovskite CsMnBr 3}. \emph{RSC Advances} \textbf{2024}, \emph{14}, 10219--10228\relax
\mciteBstWouldAddEndPuncttrue
\mciteSetBstMidEndSepPunct{\mcitedefaultmidpunct}
{\mcitedefaultendpunct}{\mcitedefaultseppunct}\relax
\EndOfBibitem
\bibitem[Rahaman and {Akther Hossain}(2018)Rahaman, and {Akther Hossain}]{Rahaman2018f}
Rahaman,~M.~Z.; {Akther Hossain},~A.~K. {Effect of metal doping on the visible light absorption, electronic structure and mechanical properties of non-toxic metal halide CsGeCl 3}. \emph{RSC Advances} \textbf{2018}, \emph{8}, 33010--33018\relax
\mciteBstWouldAddEndPuncttrue
\mciteSetBstMidEndSepPunct{\mcitedefaultmidpunct}
{\mcitedefaultendpunct}{\mcitedefaultseppunct}\relax
\EndOfBibitem
\bibitem[Krylov(2024)]{Krylov2024}
Krylov,~V.~V. {On the Applicability of Kramers–Kronig Dispersion Relations to Guided and Surface Waves}. \emph{Acoustics} \textbf{2024}, \emph{6}, 610--619\relax
\mciteBstWouldAddEndPuncttrue
\mciteSetBstMidEndSepPunct{\mcitedefaultmidpunct}
{\mcitedefaultendpunct}{\mcitedefaultseppunct}\relax
\EndOfBibitem
\bibitem[Zosiamliana \latin{et~al.}(2022)Zosiamliana, Lalrinkima, Chettri, Abdurakhmanov, Ghimire, and Rai]{Zosiamliana2022n}
Zosiamliana,~R.; Lalrinkima; Chettri,~B.; Abdurakhmanov,~G.; Ghimire,~M.~P.; Rai,~D.~P. {Electronic, mechanical, optical and piezoelectric properties of glass-like sodium silicate (Na 2 SiO 3 ) under compressive pressure}. \emph{RSC Advances} \textbf{2022}, \emph{12}, 12453--12462\relax
\mciteBstWouldAddEndPuncttrue
\mciteSetBstMidEndSepPunct{\mcitedefaultmidpunct}
{\mcitedefaultendpunct}{\mcitedefaultseppunct}\relax
\EndOfBibitem
\bibitem[Singh(2002)]{Singh2002}
Singh,~S. {Refractive Index Measurement and its Applications}. \emph{Physica Scripta} \textbf{2002}, \emph{65}, 167--180\relax
\mciteBstWouldAddEndPuncttrue
\mciteSetBstMidEndSepPunct{\mcitedefaultmidpunct}
{\mcitedefaultendpunct}{\mcitedefaultseppunct}\relax
\EndOfBibitem
\bibitem[Weaver(1976)]{Weaver1976a}
Weaver,~J. {Application of finite strain theory to non-cubic crystals}. \emph{Journal of Physics and Chemistry of Solids} \textbf{1976}, \emph{37}, 711--718\relax
\mciteBstWouldAddEndPuncttrue
\mciteSetBstMidEndSepPunct{\mcitedefaultmidpunct}
{\mcitedefaultendpunct}{\mcitedefaultseppunct}\relax
\EndOfBibitem
\bibitem[Mouhat and Coudert(2014)Mouhat, and Coudert]{Mouhat2014}
Mouhat,~F.; Coudert,~F.-X. {Necessary and sufficient elastic stability conditions in various crystal systems}. \emph{Physical Review B} \textbf{2014}, \emph{90}, 224104\relax
\mciteBstWouldAddEndPuncttrue
\mciteSetBstMidEndSepPunct{\mcitedefaultmidpunct}
{\mcitedefaultendpunct}{\mcitedefaultseppunct}\relax
\EndOfBibitem
\bibitem[Born(1940)]{Born1940d}
Born,~M. {On the stability of crystal lattices. I}. \emph{Mathematical Proceedings of the Cambridge Philosophical Society} \textbf{1940}, \emph{36}, 160--172\relax
\mciteBstWouldAddEndPuncttrue
\mciteSetBstMidEndSepPunct{\mcitedefaultmidpunct}
{\mcitedefaultendpunct}{\mcitedefaultseppunct}\relax
\EndOfBibitem
\bibitem[Isotta \latin{et~al.}(2023)Isotta, Peng, Balodhi, and Zevalkink]{Isotta2023}
Isotta,~E.; Peng,~W.; Balodhi,~A.; Zevalkink,~A. {Elastic Moduli: a Tool for Understanding Chemical Bonding and Thermal Transport in Thermoelectric Materials}. \emph{Angewandte Chemie International Edition} \textbf{2023}, \emph{62}, e202213649\relax
\mciteBstWouldAddEndPuncttrue
\mciteSetBstMidEndSepPunct{\mcitedefaultmidpunct}
{\mcitedefaultendpunct}{\mcitedefaultseppunct}\relax
\EndOfBibitem
\bibitem[Kleinman(1962)]{Kleinman1962b}
Kleinman,~L. {Deformation Potentials in Silicon. I. Uniaxial Strain}. \emph{Physical Review} \textbf{1962}, \emph{128}, 2614--2621\relax
\mciteBstWouldAddEndPuncttrue
\mciteSetBstMidEndSepPunct{\mcitedefaultmidpunct}
{\mcitedefaultendpunct}{\mcitedefaultseppunct}\relax
\EndOfBibitem
\bibitem[Ahmed \latin{et~al.}(2023)Ahmed, Mahamudujjaman, Afzal, Islam, Islam, and Naqib]{Ahmed2023b}
Ahmed,~R.; Mahamudujjaman,~M.; Afzal,~M.~A.; Islam,~M.~S.; Islam,~R.; Naqib,~S. {DFT based comparative analysis of the physical properties of some binary transition metal carbides XC (X = Nb, Ta, Ti)}. \emph{Journal of Materials Research and Technology} \textbf{2023}, \emph{24}, 4808--4832\relax
\mciteBstWouldAddEndPuncttrue
\mciteSetBstMidEndSepPunct{\mcitedefaultmidpunct}
{\mcitedefaultendpunct}{\mcitedefaultseppunct}\relax
\EndOfBibitem
\bibitem[Dovale-Farelo \latin{et~al.}(2022)Dovale-Farelo, Tavadze, Lang, Bautista-Hernandez, and Romero]{Dovale-Farelo2022a}
Dovale-Farelo,~V.; Tavadze,~P.; Lang,~L.; Bautista-Hernandez,~A.; Romero,~A.~H. {Vickers hardness prediction from machine learning methods}. \emph{Scientific Reports} \textbf{2022}, \emph{12}, 22475\relax
\mciteBstWouldAddEndPuncttrue
\mciteSetBstMidEndSepPunct{\mcitedefaultmidpunct}
{\mcitedefaultendpunct}{\mcitedefaultseppunct}\relax
\EndOfBibitem
\bibitem[Singh \latin{et~al.}(2023)Singh, Kaur, Shandilya, Rana, Rai, Mishra, Syv{\"{a}}j{\"{a}}rvi, and Tiwari]{Singh2023c}
Singh,~P.; Kaur,~G.; Shandilya,~M.; Rana,~P.; Rai,~R.; Mishra,~Y.; Syv{\"{a}}j{\"{a}}rvi,~M.; Tiwari,~A. {Trends in piezoelectric nanomaterials towards green energy scavenging nanodevices}. \emph{Materials Today Sustainability} \textbf{2023}, \emph{24}, 100583\relax
\mciteBstWouldAddEndPuncttrue
\mciteSetBstMidEndSepPunct{\mcitedefaultmidpunct}
{\mcitedefaultendpunct}{\mcitedefaultseppunct}\relax
\EndOfBibitem
\bibitem[Li and Lee(2022)Li, and Lee]{Li2022b}
Li,~T.; Lee,~P.~S. {Piezoelectric Energy Harvesting Technology: From Materials, Structures, to Applications}. \emph{Small Structures} \textbf{2022}, \emph{3}, 2100128\relax
\mciteBstWouldAddEndPuncttrue
\mciteSetBstMidEndSepPunct{\mcitedefaultmidpunct}
{\mcitedefaultendpunct}{\mcitedefaultseppunct}\relax
\EndOfBibitem
\bibitem[King-Smith and Vanderbilt(1993)King-Smith, and Vanderbilt]{King-Smith1993c}
King-Smith,~R.~D.; Vanderbilt,~D. {Theory of polarization of crystalline solids}. \emph{Physical Review B} \textbf{1993}, \emph{47}, 1651--1654\relax
\mciteBstWouldAddEndPuncttrue
\mciteSetBstMidEndSepPunct{\mcitedefaultmidpunct}
{\mcitedefaultendpunct}{\mcitedefaultseppunct}\relax
\EndOfBibitem
\bibitem[Saigusa(2017)]{Saigusa2017}
Saigusa,~Y. \emph{Advanced Piezoelectric Materials}; Elsevier, 2017; pp 197--233\relax
\mciteBstWouldAddEndPuncttrue
\mciteSetBstMidEndSepPunct{\mcitedefaultmidpunct}
{\mcitedefaultendpunct}{\mcitedefaultseppunct}\relax
\EndOfBibitem
\bibitem[Valasek(1922)]{Valasek1922}
Valasek,~J. {Properties of Rochelle Salt Related to the Piezo-electric Effect}. \emph{Physical Review} \textbf{1922}, \emph{20}, 639--664\relax
\mciteBstWouldAddEndPuncttrue
\mciteSetBstMidEndSepPunct{\mcitedefaultmidpunct}
{\mcitedefaultendpunct}{\mcitedefaultseppunct}\relax
\EndOfBibitem
\bibitem[Bairagi \latin{et~al.}(2023)Bairagi, ul~Islam, Shahadat, Mulvihill, and Ali]{Bairagi2023}
Bairagi,~S.; ul~Islam,~S.; Shahadat,~M.; Mulvihill,~D.~M.; Ali,~W. {Mechanical energy harvesting and self-powered electronic applications of textile-based piezoelectric nanogenerators: A systematic review}. \emph{Nano Energy} \textbf{2023}, \emph{111}, 108414\relax
\mciteBstWouldAddEndPuncttrue
\mciteSetBstMidEndSepPunct{\mcitedefaultmidpunct}
{\mcitedefaultendpunct}{\mcitedefaultseppunct}\relax
\EndOfBibitem
\bibitem[Panda(2009)]{Panda2009}
Panda,~P.~K. {Review: environmental friendly lead-free piezoelectric materials}. \emph{Journal of Materials Science} \textbf{2009}, \emph{44}, 5049--5062\relax
\mciteBstWouldAddEndPuncttrue
\mciteSetBstMidEndSepPunct{\mcitedefaultmidpunct}
{\mcitedefaultendpunct}{\mcitedefaultseppunct}\relax
\EndOfBibitem
\bibitem[Bassiri-Gharb(2008)]{Bassiri-Gharb2008a}
Bassiri-Gharb,~N. \emph{Piezoelectric and Acoustic Materials for Transducer Applications}; Springer US: Boston, MA, 2008; pp 413--430\relax
\mciteBstWouldAddEndPuncttrue
\mciteSetBstMidEndSepPunct{\mcitedefaultmidpunct}
{\mcitedefaultendpunct}{\mcitedefaultseppunct}\relax
\EndOfBibitem
\bibitem[Bernardini \latin{et~al.}(1997)Bernardini, Fiorentini, and Vanderbilt]{Bernardini1997c}
Bernardini,~F.; Fiorentini,~V.; Vanderbilt,~D. {Spontaneous polarization and piezoelectric constants of III-V nitrides}. \emph{Physical Review B} \textbf{1997}, \emph{56}, R10024--R10027\relax
\mciteBstWouldAddEndPuncttrue
\mciteSetBstMidEndSepPunct{\mcitedefaultmidpunct}
{\mcitedefaultendpunct}{\mcitedefaultseppunct}\relax
\EndOfBibitem
\bibitem[Rohrlich(2009)]{Rohrlich2009c}
Rohrlich,~D. \emph{Compendium of Quantum Physics}; Springer Berlin Heidelberg: Berlin, Heidelberg, 2009; pp 31--36\relax
\mciteBstWouldAddEndPuncttrue
\mciteSetBstMidEndSepPunct{\mcitedefaultmidpunct}
{\mcitedefaultendpunct}{\mcitedefaultseppunct}\relax
\EndOfBibitem
\bibitem[Bechmann(1958)]{Bechmann1958d}
Bechmann,~R. {Elastic and Piezoelectric Constants of Alpha-Quartz}. \emph{Physical Review} \textbf{1958}, \emph{110}, 1060--1061\relax
\mciteBstWouldAddEndPuncttrue
\mciteSetBstMidEndSepPunct{\mcitedefaultmidpunct}
{\mcitedefaultendpunct}{\mcitedefaultseppunct}\relax
\EndOfBibitem
\bibitem[Tarumi \latin{et~al.}(2007)Tarumi, Nakamura, Ogi, and Hirao]{Tarumi2007c}
Tarumi,~R.; Nakamura,~K.; Ogi,~H.; Hirao,~M. {Complete set of elastic and piezoelectric coefficients of $\alpha$-quartz at low temperatures}. \emph{Journal of Applied Physics} \textbf{2007}, \emph{102}\relax
\mciteBstWouldAddEndPuncttrue
\mciteSetBstMidEndSepPunct{\mcitedefaultmidpunct}
{\mcitedefaultendpunct}{\mcitedefaultseppunct}\relax
\EndOfBibitem
\bibitem[Zosiamliana \latin{et~al.}(2022)Zosiamliana, Lalrinkima, Chettri, Abdurakhmanov, Ghimire, and Rai]{Zosiamliana2022o}
Zosiamliana,~R.; Lalrinkima; Chettri,~B.; Abdurakhmanov,~G.; Ghimire,~M.~P.; Rai,~D.~P. {Electronic, mechanical, optical and piezoelectric properties of glass-like sodium silicate (Na 2 SiO 3 ) under compressive pressure}. \emph{RSC Advances} \textbf{2022}, \emph{12}, 12453--12462\relax
\mciteBstWouldAddEndPuncttrue
\mciteSetBstMidEndSepPunct{\mcitedefaultmidpunct}
{\mcitedefaultendpunct}{\mcitedefaultseppunct}\relax
\EndOfBibitem
\bibitem[Chen \latin{et~al.}(2019)Chen, Kong, Song, Jiang, Tian, Yu, Qin, Wang, and Zhao]{Chen2019}
Chen,~F.; Kong,~L.; Song,~W.; Jiang,~C.; Tian,~S.; Yu,~F.; Qin,~L.; Wang,~C.; Zhao,~X. The electromechanical features of LiNbO3 crystal for potential high temperature piezoelectric applications. \emph{Journal of Materiomics} \textbf{2019}, \emph{5}, 73--80\relax
\mciteBstWouldAddEndPuncttrue
\mciteSetBstMidEndSepPunct{\mcitedefaultmidpunct}
{\mcitedefaultendpunct}{\mcitedefaultseppunct}\relax
\EndOfBibitem
\bibitem[Fu \latin{et~al.}(2007)Fu, Liu, Cheng, Bhalla, and Guo]{Fu2007}
Fu,~J.~Y.; Liu,~P.~Y.; Cheng,~J.; Bhalla,~A.~S.; Guo,~R. Optical measurement of the converse piezoelectric d33 coefficients of bulk and microtubular zinc oxide crystals. \emph{Applied Physics Letters} \textbf{2007}, \emph{90}, 212907\relax
\mciteBstWouldAddEndPuncttrue
\mciteSetBstMidEndSepPunct{\mcitedefaultmidpunct}
{\mcitedefaultendpunct}{\mcitedefaultseppunct}\relax
\EndOfBibitem
\bibitem[Zhao \latin{et~al.}(2018)Zhao, Ma, Liu, Yu, and Cai]{Zhao2018}
Zhao,~Y.-Q.; Ma,~Q.-R.; Liu,~B.; Yu,~Z.-L.; Cai,~M.-Q. Pressure-induced strong ferroelectric polarization in tetra-phase perovskite CsPbBr3. \emph{Phys. Chem. Chem. Phys.} \textbf{2018}, \emph{20}, 14718--14724\relax
\mciteBstWouldAddEndPuncttrue
\mciteSetBstMidEndSepPunct{\mcitedefaultmidpunct}
{\mcitedefaultendpunct}{\mcitedefaultseppunct}\relax
\EndOfBibitem
\bibitem[Roy \latin{et~al.}(2012)Roy, Bennett, Rabe, and Vanderbilt]{Roy2012d}
Roy,~A.; Bennett,~J.~W.; Rabe,~K.~M.; Vanderbilt,~D. {Half-Heusler Semiconductors as Piezoelectrics}. \emph{Physical Review Letters} \textbf{2012}, \emph{109}, 037602\relax
\mciteBstWouldAddEndPuncttrue
\mciteSetBstMidEndSepPunct{\mcitedefaultmidpunct}
{\mcitedefaultendpunct}{\mcitedefaultseppunct}\relax
\EndOfBibitem
\bibitem[Otto(1992)]{Otto1992b}
Otto,~P. {Calculation of the polarizability and hyperpolarizabilities of periodic quasi-one-dimensional systems}. \emph{Physical Review B} \textbf{1992}, \emph{45}, 10876--10885\relax
\mciteBstWouldAddEndPuncttrue
\mciteSetBstMidEndSepPunct{\mcitedefaultmidpunct}
{\mcitedefaultendpunct}{\mcitedefaultseppunct}\relax
\EndOfBibitem
\end{mcitethebibliography}

\end{document}